\documentclass[prc,aps,amsmath,amssymb,superscriptaddress,twocolumn,showpacs,floatfix,a4paper]{revtex4-2}

\usepackage{graphicx,colordvi}
\usepackage{dcolumn}
\usepackage{bm}
\usepackage{threeparttable}
\usepackage{xspace}
\usepackage{booktabs}
\usepackage{gensymb}
\usepackage{cases}
\usepackage{textcomp}
\usepackage{tabularx,booktabs}
\usepackage{epstopdf}
\usepackage{physics} 
\usepackage{mathtools, amssymb, amsthm, amsmath}

\usepackage{appendix}
\usepackage{csquotes}

\usepackage{xcolor}
\usepackage{tcolorbox}

\usepackage[dvipsnames]{xcolor}
\usepackage{bigints}
\usepackage{scalerel}

\begin{document}

\title{Total angular momentum distributions of fission fragments at scission in a 2-center basis}

\author{P. Carpentier}
\affiliation{%
  Universit\'e Paris-Saclay, CEA, LMCE, 91680, Bruyères-le-Châtel, France
}%

\author{N. Pillet}%
\affiliation{%
  Universit\'e Paris-Saclay, CEA, LMCE, 91680, Bruyères-le-Châtel, France
}%
\affiliation{%
 CEA, DAM, DIF, F-91297 Arpajon cedex, France
}%

\author{P. Nieto Gallego}%
\affiliation{%
  Universit\'e Paris-Saclay, CEA, LMCE, 91680, Bruyères-le-Châtel, France
}%
\affiliation{%
 CEA, DAM, DIF, F-91297 Arpajon cedex, France
}%

\date{\today}

\begin{abstract}
In this article, we propose a new method to calculate the angular momentum distribution of fission fragments adapted to the case of 2-center axial basis used to solve the Hartree-Fock-Bogoliubov equations in view of fission studies. As a first application, we have studied the properties of the fission fragment angular momentum distribution in the scission area of $^{240}$Pu using the Gogny D1S interaction. Here, a one dimensional asymmetric and adiabatic path (the lowest in energy), generated by the newly proposed Link and Drop methods \cite{carpentier2024,trilogy1}, has been considered. The major result of this study is the appearance of an odd-even angular momentum staggering in both fragments when going through scission. The odd-$J$ components are progressively suppressed as the fragments separate, leading to an increasing dominance of even-$J$ components. This behavior is systematically correlated with the rapid reduction of the intrinsic octupole deformation of the fragments, associated with the restoration of a more nearly reflection-symmetric shape. Finally, particle-number projection reveals the role of shell effects associated with specific fragment configurations in shaping the fragment angular momentum distributions.

\end{abstract}

\maketitle

\section{Introduction}

Nuclear fission is one of the most fascinating phenomenon 
during which a nucleus separates most likely into two 
fragments, that de-excite themselves by emitting prompt 
neutrons and/or gammas. The complexity of this dynamical 
process implying heavy nuclei comes from the interplay 
between collective and intrinsic degrees of freedom. These 
different aspects are reflected in various observables now 
experimentally accessible in many systems, such as fragment 
mass and charge yields, total kinetic energy (TKE), total 
excitation energy (TXE), prompt neutron and $\gamma$-ray 
multiplicities or angular momentum (AM). 
While shell effects are one of the major understanding keys 
(see \cite{morfou,scamps2018,Schmitt2021}, for example), many newly proposed experiments 
\cite{ramos2019,ramos2020,ramos2026,wilson2021,schmidt2000,schmidt2021} and 
theoretical models, both phenomenological and microscopic 
\cite{geff,SiSi,Seki,SCAMPS,RBUL,bulgac2023,OldVret,hillwheeler1,griffinwheeler1,TDGCM3,bergerTDGCM,DynaGou,
younes1,DynaRe,LastVret}, tend to 
study the correlations between fission observables in order 
to elucidate the origin of fission fragment 
properties at scission. To make things a little more 
complicated, the distributions of these observables 
depend on the initial conditions of the reaction.

Among the intrinsic properties, the AM of fission fragments has manifested a renewed interest this last years on both experimental and theoretical sides. It is known from a long time that fission fragments are generally produced with angular momenta reaching several units of $\hbar$, while fissioning nuclei can display a small initial one. To explain this AM generation, several mechanisms have been proposed in the past as well as more recently, including the excitation of collective modes before scission, bending, twisting and wriggling motions, fragment deformations, neck dynamics as well as torques acting during the separation of the fragments \cite{rasmussen1969,dietrich1974,mikhailov1999,mikhailov2001,scamps2023,scamps2023b}. 
The intrinsic AM of fission fragments is not directly accessible experimentally at scission. Nowaday, it can be studied from their prompt $\gamma$ emission or the population of isomeric states. 
In this perspective, a major experiment has been achieved 
by Wilson et \textit{al.} in several fissioning systems 
\cite{wilson2021} which has revealed the absence of a 
strong correlation between the spins of complementary 
fragments. This result led the authors to the conclusion 
that part of the AM may be generated after 
scission. This interpretation stimulated further 
theoretical investigations showing that pre-scission 
mechanisms can also produce weakly correlated spin 
distributions between fragments 
\cite{randrup2014,vogt2025,nicholson2021}, while some 
microscopic analyzes \cite{scamps2023,scamps2023b} 
supported these experimental observations. 
Since then, several experiments have been conducted to supplement the relevant observables and identify correlations among them. 

Several experimental approaches are now available to constrain the AM distributions of fission fragments. Nicholson and  collaborators have investigated isomeric state population ratios and the corresponding angular momenta of $^{132}\mathrm{Sn}$ populated in the $^{241}\mathrm{Pu}(n_{\mathrm{th}},f)$ reaction \cite{nicholson2021}.
Indeed, although it is an indirect measurement, the probability of populating a metastable state depends on the spin distribution of the fragment.
Giha et \textit{al.} have studied correlations between fragment energy and $\gamma$-ray emission in $^{239}\mathrm{Pu}(n,f)$ reaction, providing constraints on how the available excitation energy is distributed among the different degrees of freedom \cite{giha2023}. A further step has been achieved with the measurement of the spin of $^{144}\mathrm{Ba}$ as a function of the TKE in the spontaneous fission of $^{252}\mathrm{Cf}$ \cite{giha2025}.
In the same time, Cannarozzo et \textit{al.} have investigated the respective contributions of excitation energy and initial compound-nucleus AM to the AM of the fission fragments \cite{cannarozzo2025}. Indeed, as already invoked in the old theoretical study \cite{dietrich1974}, the final AM distribution of the fragments depend not only on the scission dynamics but also on the AM and excitation energy of the compound nucleus. 
Finally, the work of Francheteau et \textit{al.} has revealed the strong correlation between the deformation of fragments resulting from the rare cold fission events 
(neutronless), their excitation energy, and their initial AM \cite{francheteau2024,francheteau2025,francheteau2026}. 
By combining precise measurements of scission deformation (e.g. in the fragmentation $^{120}$Cd/$^{132}$Sn) with a prompt $\gamma$-ray spectrum analysis, they demonstrate that the AM of the fragments is directly linked to their shape and energy at the moment of separation. Using high-resolution detectors and genetic algorithms, this study has been extended to other fragmentations like the $^{118}$Pd/$^{134}$Te one, providing a unified understanding of correlated observables in the context of cold fission.\\

As previously mentioned, an ultimate goal of microscopic approaches is to develop a description from which the dynamical scission configurations allows to determine the properties of the fragments consistently.
Two main families of approaches are developed to reach this objective, namely the Time-Dependent 
Hartree-Fock-Bogoliubov approach (TDHFB) \cite{SiSi,Seki,SCAMPS,RBUL,bulgac2023,OldVret} and the Time-Dependent Generator Coordinate 
Method (TDGCM) \cite{younes1,TDGCM3,bergerTDGCM,DynaGou,
younes1,DynaRe}. While the first one takes into account one-body dissipative effects, 
the second one introduces naturally the collective deformations responsible for the yield distribution 
describing symmetric to very asymmetric fragmentations. In order to take advantage of both 
approaches, there also exists some attempts to combine them \cite{LastVret}.

The present work contributes to this broader objective in the context of the recently developed Schrödinger Collective-Intrinsic Model (SCIM) \cite{TPaul,trilogy1,trilogy2,trilogy3,bernard1,TBernard}. The central idea of the SCIM is to extend the adiabatic TDGCM wave-function ansatz by incorporating intrinsic excited states, thereby providing a framework in which diabatic effects can be taken into account. Its first implementation relies on Hartree-Fock-Bogoliubov (HFB) calculations based on recently introduced overlap constraints \cite{carpentier2024,trilogy1}, defining the \enquote{Link}, \enquote{Drop}, and \enquote{Continuous deflation} methods. These methods make it possible to construct continuous and regular one-dimensional adiabatic and excited fission paths extending from the inner barrier region through scission and beyond. In the scission region, the use of a 2-center \cite{TBerger} basis facilitates the disentanglement of the two emerging fragments. In contrast, within the standard adiabatic TDGCM framework, using the Gaussian overlap approximation and without overlap constraints, fragment properties are generally extracted while a finite number of particles remains in the neck \cite{marevic2021,marevic2025,marevic2026}. The first application of SCIM to $^{240}$Pu demonstrated the ability of this framework to extract several fragment observables at scission, including a complete energy balance involving both the total kinetic energy (TKE) and total excitation energy (TXE). The fragment AM distributions, however, could not be evaluated because the use of a 2-center basis requires specific developments beyond those needed for a standard 1-center basis.

To complete the description of fission-fragment properties at scission, a new method is developed to calculate the total AM distributions of fission fragments for HFB states expanded in a 2-center basis. The formalism developed here provides a general framework for calculating these distributions from HFB states expanded in either 1-center or 2-center bases. While the 1-center case leads to relatively simple spatial matrix elements of the single-particle rotation operator, the 2-center case presents an additional challenge: the translations of the basis functions along the symmetry axis become coupled to the rotation, resulting in a non-trivial analytical problem. The resulting formulation provides an analytical treatment of this coupling and leads to an efficient numerical implementation based on multivariate Hermite polynomials.

The article is organized as follows. Section \ref{theory} introduces the general framework for total AM projection in a 2-center basis. In subsection \ref{sub1ME}, the matrix elements of the single-particle rotation operator are derived using the properties of generating functions and multivariate Hermite polynomials. Subsection \ref{sub2ROTA}, \ref{sub3QUAD} and \ref{sub4PN} are devoted to the calculation of the total AM projected norm and presents the extension of the formalism to particle number projection. Section \ref{section4} then presents applications to the scission region of $^{240}$Pu along an asymmetric adiabatic fission path obtained with the recently proposed Link and Drop methods \cite{carpentier2024,trilogy1}. The resulting fragment total AM distributions are analyzed, with particular emphasis on the emergence of an odd-even staggering and its relation to the reduction of the intrinsic octupole deformation of the fragments. Particle number projection is further used to resolve the total AM distributions according to the different fragmentations contained in the HFB states and to investigate the role of shell effects. Finally, section \ref{section6} summarizes the main results and discusses perspectives for future developments.

\section{Theoretical framework}\label{theory}

In general, the projection operator onto a state of total AM $J$ can be written as
\begin{align}
    \hat P^{J}_{MK} = \frac{2J+1}{8 \pi^2}\int d\Omega\, D^{J*}_{MK}(\Omega)\hat R(\Omega),
\end{align}
where $J$ is the total AM, $M$ and $K$ are its projections onto the $z$ axis of the laboratory frame and the $z'$ axis of the intrinsic frame, respectively, $\Omega$ denotes the Euler angles, $\hat R(\Omega)$ is the rotation operator, and $D^{J}_{MK}(\Omega)$ are the Wigner matrices.

In this work, we consider the total AM projection of HFB states expanded in a 2-center axially symmetric harmonic oscillator (HO) basis \cite{TBerger,TPaul,trilogy1}. This choice of basis is particularly well suited to the description of the strongly elongated configurations encountered along the fission path and provides a natural framework for describing the two nascent fragments. However, its 2-center character introduces additional theoretical and computational difficulties. In particular, the translation inherent to the 2-center basis and the rotation operation are coupled in a non-trivial way, making the construction of the matrix elements of the single-particle rotation operator more involved than in a standard 1-center HO basis. To the best of our knowledge, these specific issues have not been addressed previously. Their consistent treatment is therefore a central aspect of the present approach.

Exploiting rotational invariance, the magnetic quantum number $M$ is degenerate, and the AM distribution can therefore be obtained from a single $M$ component. For an axially symmetric intrinsic state with $K=0$, we choose $M=0$ without loss of generality. The projector associated with a given total AM $J$ then reduces to:
\begin{align}
 \hat P^{J} = \frac{2J + 1}{2}
 \int_0^\pi d\zeta\, \sin(\zeta)\,
 P_J\left(\cos\zeta\right)\hat R_y(\zeta),
\end{align}
where $\hat R_y(\zeta)$ denotes a rotation by an angle $\zeta$ about the $y$ axis and $P_J$ is the Legendre polynomial of order $J$. The probability $P(J)$ of finding the intrinsic state in a state of total AM $J$ is consequently given by:
\begin{align}\label{eqPJ}
    P(J) = \frac{2J + 1}{2}
    \int_0^\pi d\zeta\, \sin(\zeta)\,
    P_J\left(\cos\zeta\right)
    \langle \Phi|\Phi(\zeta)\rangle,
\end{align}
where
\begin{align}
    \langle \Phi|\Phi(\zeta)\rangle
    =
    \langle \Phi|\hat R_y(\zeta)|\Phi\rangle
\end{align}
is the rotated overlap kernel.

The calculation of the probability distribution $P(J)$ is organized into three steps. First, the action of the rotation operator on the single-particle states is determined and the corresponding matrix elements are evaluated. These matrix elements are then used to construct the rotated overlap kernel. The angular integration is subsequently performed using an appropriate numerical quadrature. 

Finally, the extension of the formalism to the simultaneous projection onto total AM and particle number is discussed.

\subsection{Matrix elements of the single-particle rotation operator}\label{sub1ME}

We consider a 2-center axial HO basis, for which the single-particle states $|\nu\rangle$ can be written as a direct product of a spatial and intrinsic spin parts:
\begin{eqnarray}
|\nu\rangle
&=&
|\nu_{\vec{r}}\rangle
\otimes |\nu_s\rangle, \nonumber \\ 
&=&
|(n_\perp,m,n_z),(b_z,b_r,d)\rangle
\otimes |s\rangle.
\end{eqnarray}
Here, $d$ denotes the displacement of the HO center along the $z$-axis, while $b_r$ and $b_z$ are the transverse and longitudinal oscillator lengths, respectively.
The set $\{n_\perp,m,n_z\}$ designates the quantum numbers of the axial HO states. 

The basic quantity entering the AM projection is the single-particle rotation matrix element $R^s_{\nu\nu'}(\zeta)$ defined by: 
\begin{eqnarray}
R^s_{\nu\nu'}(\zeta)
&=&
\langle \nu|
e^{\displaystyle -i\zeta\hat j_y}
|\nu'\rangle \nonumber \\
&=&
\langle \nu_{\vec{r}}|
e^{\displaystyle -i\zeta\hat l_y}
|\nu'_{\vec{r}}\rangle\langle \nu_{s}|
e^{\displaystyle -i\zeta\hat s_y}
|\nu'_{s}\rangle \nonumber \\
&=&
R_{\nu\nu'}(\zeta) ~ 
S_{s_\nu s_{\nu'}}(\zeta).
\end{eqnarray}
where $R_{\nu\nu'}(\zeta)$ and $S_{s_\nu s_{\nu'}}(\zeta)$ represent its spatial and spin contributions. In the following, we discuss their explicit calculation.

\subsubsection{Spatial contribution}

The spatial contribution $R_{\nu\nu'}(\zeta)$ to the matrix element $R^s_{\nu\nu'}(\zeta)$ is given by:
\begin{eqnarray}\label{eq3}
R_{\nu\nu'}(\zeta)=\int d\vec{r} \;
\psi_\nu^*(\vec{r})
\psi_{\nu'}\bigl(\vec{r}(\zeta)\bigr),
\end{eqnarray}
where $\vec{r}(\zeta)$ denotes the position vector obtained by rotating $\vec{r}$ about the $y$-axis. More explicitly, it is defined by applying the transformation $D(\zeta)$ such that:
\begin{eqnarray}
    \vec{r}(\zeta) &=& D(\zeta) \vec{r}
\end{eqnarray}
with
\begin{eqnarray}
     D(\zeta)=
\begin{pmatrix}
\cos\zeta&0&-\sin\zeta\\
0&1&0\\
\sin\zeta&0&\cos\zeta
\end{pmatrix}.
\end{eqnarray}

\noindent As well-known, the spatial part $\psi_\nu(\vec{r})$ of the wave functions associated with the 2-center axial HO basis can be factorized into a longitudinal part $\varphi_{n_z}(z+d,b_z) $ along the $z$-axis and a transverse part $\phi_{(m,n_\perp)}(\vec{r}_\perp,b_r)$, perpendicular to the $z$-axis characterized by the vector $\vec{r}_\perp$ in such a way that:
\begin{align}
&\psi_\nu(\vec{r}) =  \varphi_{n_z}(z+d,b_z) \times  \phi_{(m,n_\perp)}(\vec{r}_\perp,b_r).  
\end{align}
\noindent The expression of the function $\varphi_{n_z}\left(z,b_z\right)$ in terms of the Hermite polynomials $H_{n_z}$ is given by:
\begin{align}
\varphi_{n_z}\left(z,b_z\right) = \frac{1}{\sqrt{2^{n_z}b_z n_z!\sqrt{\pi}}} ~ e^{\displaystyle -\frac{1}{2}\left(\frac{z}{b_z}\right)^2} H_{n_z} \left(\frac{z}{b_z} \right).
\end{align}

\noindent The transverse wave function $\phi_{(m,n_\perp)}\left(\vec{r}_\perp,b_r \right)$ can be expressed in terms of the associated Laguerre polynomials $L_{n_\perp}^{|m|}$:
\begin{eqnarray}
\phi_{(m,n_\perp)}\left(\vec{r}_\perp,b_r \right) &=& \frac{1}{b_r\sqrt{\pi}}\sqrt{\frac{n_\perp !}{(n_\perp + |m|)!}}e^{\displaystyle im\phi}\left(\frac{r_\perp}{b_r}\right)^{\displaystyle |m|} \nonumber \\ && \times L_{n_\perp}^{|m|}\left[(\frac{r_\perp}{b_r})^2 \right]e^{\displaystyle -\frac{1}{2}\left(\frac{r_\perp}{b_r}\right)^{2}}. 
\end{eqnarray}

\noindent The evaluation of the spatial matrix elements is particularly challenging because the 2-center basis involves both translations along the symmetry axis and rotations, which become non-trivially coupled in the resulting integrals \eqref{eq3}. To address this difficulty, we have developed an approach based on generating functions and multivariate Hermite polynomials.
The key idea is to represent the HO wave functions in terms of Gaussian generating functions. Then, the resulting spatial integrals can be evaluated analytically using standard Gaussian integration techniques. The multivariate Hermite polynomials are subsequently used to extract the desired spatial matrix elements.\\

Our primary goal is to determine the Gaussian representation of the HO wave functions.
The longitudinal wave functions are treated using their generating function. For a given value of $d$, it can be shown that:
\begin{eqnarray}
   e^{\displaystyle \left[-\frac{1}{2}(\frac{z+d}{b_z}
   )^2-p_z^2+2p_z\frac{z+d}{b_z}\right]} &=& \sqrt{b_z\sqrt{\pi}} \nonumber \\  \times\sum_{n_z=0}^{+\infty} p_z^{n_z}\sqrt{\frac{2^{n_z}}{n_z!}}\varphi_{n_z}(z+d,b_z).
\end{eqnarray}
where $p_z$ is the generating variable.\\

\noindent As a consequence, the wave function $\varphi_{n_z}(z+d,b_z)$ can be represented as a derivative with respect to $p_z$:
\begin{align}
&\varphi_{n_z}(z+d,b_z)
=
\mathcal{N}_{n_z} 
\nonumber \\ &\times \left.
\frac{\partial^{n_z}}{\partial p_z^{n_z}} 
 ~ e^{\displaystyle \left[
-p_z^2
+2p_z\frac{z+d}{b_z}
-\frac{z^2+2zd}{2b_z^2}
\right]}
\right|_{p_z=0}
\end{align}
\noindent with the following expression for the coefficient $\mathcal{N}_{n_z}$:
\begin{eqnarray}
   \mathcal{N}_{n_z} &=& \frac{\displaystyle e^{\displaystyle \frac{-d^2}{2b_z^2}} }{ \displaystyle \sqrt{2^{n_z} n_z !b_z\sqrt{\pi}}}.
\end{eqnarray}

\noindent The transverse part of the wave function is conveniently expressed in terms of the complex variables $p_\pm$ associated with the generating variable $p_{\perp}$:
\begin{align}
   &\displaystyle e^{\displaystyle [-\frac{1}{2}(\frac{r_\perp}{b_r})^2+\frac{2}{b_r}\vec{r}_\perp.\vec{p}_\perp-p_\perp^2]} = b_r\sqrt{\pi} \\ & \displaystyle \qquad \qquad  \times\sum_{n_+,n_-}\frac{\displaystyle (-1)^{\text{min}(n_+,n_-)}\sqrt{2^{n_+ +n_-}}}{\sqrt{n_-!n_+!}} \nonumber \\ & \displaystyle \qquad \qquad \times ~p_+^{n_+}~p_-^{n_-} ~ \phi_{(n_+,n_-)}(\vec{r}_\perp,b_r) \nonumber 
\end{align}
with 
\begin{eqnarray}
   p_- = \frac{\displaystyle p_\perp}{\sqrt{2}} e^{\displaystyle -i\varphi_p}~, \\ p_+ = \frac{\displaystyle p_\perp}{\sqrt{2}} e^{\displaystyle +i\varphi_p} ~, 
\end{eqnarray}
and
\begin{align}
    &n_+ = \begin{cases} n_\perp +m \qquad \text{if $m \geq 0$}\\
    n_\perp \qquad \qquad\text{if $m < 0$}
    \end{cases}, \\ \nonumber  \\ &n_- = \begin{cases} n_\perp \qquad \qquad\text{if $m \geq 0$}  \\
    n_\perp - m \qquad \text{if $m < 0$}
    \end{cases}. 
\end{align}
The transverse wave functions can likewise be obtained by taking partial derivatives with respect to the generating variables $p_+$ and $p_-$:
\begin{align}
   &\phi_{(n_+,n_-)}(\vec{r}_\perp,b_r) = \mathcal{N}_{n_+n_-}\\& \times \left.\frac{\displaystyle \partial^{n_+}\partial^{n_-}}{\partial p_+^{n_+}\partial p_-^{n_-}}~ e^{[\displaystyle -\frac{1}{2}(\frac{r_\perp}{b_r})^2+\frac{2}{b_r}\vec{r}_\perp.\vec{p}_\perp-p_\perp^2]}\right|_{p_+=p_-=0}, \nonumber 
\end{align}
 where the coefficients $\mathcal{N}_{n_+n_-}$ have for expression:
\begin{align}
    &\mathcal{N}_{n_+n_-} = \frac{\displaystyle 1}{\displaystyle b_r\sqrt{\pi} (-1)^{\text{min}(n_+,n_-)}\sqrt{2^{n_+ +n_-}n_+!n_-!}}. 
\end{align}
\noindent It should also be noted that:
\begin{align}
    \phi_{(n_+,n_-)}^*(\vec{r}_\perp,b_r) = \phi_{(n_-,n_+)}(\vec{r}_\perp,b_r).
\end{align}

\noindent Using the Gaussian representations of the longitudinal and transverse wave functions derived previously, the expression of the spatial matrix element $R_{\nu\nu'}\left( \zeta \right)$ becomes:
\begin{widetext}
\begin{align}
    & \displaystyle  R_{\nu\nu'}\left( \zeta \right) = \mathcal{N}_{n_z}\mathcal{N}_{n_-n_+}\mathcal{N}_{n_z'}\mathcal{N}_{n_+'n_-'} ~ 
    \frac{\displaystyle \partial^{n_z}\partial^{n_-}\partial^{n_+}}{\partial p_z^{n_z}\partial p_+^{n_-}\partial p_-^{n_+}} ~   \frac{\displaystyle \partial^{n'_z}\partial^{n'_+}\partial^{'n_-}}{\partial q_z^{n'_z}\partial q_+^{n'_+}\partial q_-^{n'_-}} ~e^{\displaystyle \left[-p^2 - q^2+\frac{\displaystyle2p_zd+2q_zd'}{b_z}\right]} \\ & \displaystyle 
 \left. \int \; d\vec{r} \;
     e^{\displaystyle \left[\frac{\displaystyle 2p_zz}{b_z}  -\frac{\displaystyle z^2+2zd}{2b_z^2} -\frac{\displaystyle 1}{2}\left(\frac{\displaystyle r_\perp}{b_r}\right)^2+\frac{2}{b_r}\vec{r}_\perp.\vec{p}_\perp + \frac{\displaystyle 2q_zz(\zeta)}{b_z}  -\frac{\displaystyle z^2(\zeta)+2z(\zeta)d'}{2b_z^2} -\frac{1}{2}\left(\frac{\displaystyle r_\perp(\zeta)}{b_r}\right)^2+\frac{\displaystyle 2}{b_r}\vec{r}_\perp(\zeta).\vec{q}_\perp \right]}\right|_{p=q=0}. \nonumber
\end{align}
\end{widetext}
In the following, we introduce the matrices $B$, $D(\zeta)$, $J(\zeta)$ and $M(\zeta)$ whose expressions are:
\begin{align}
    B = \begin{pmatrix}
       \displaystyle  \frac{1}{b_r} & 0 & 0 \\
        0 & \displaystyle \frac{1}{b_r} & 0 \\
        0 & 0 & \displaystyle \frac{1}{b_z}
    \end{pmatrix},
\end{align}
\begin{align}
   D(\zeta) = \begin{pmatrix}
        \displaystyle \text{cos}(\zeta) & 0 & \displaystyle -\text{sin}(\zeta) \\
        0 & 1 & 0 \\
      \displaystyle  \text{sin}(\zeta) & 0 & \displaystyle \text{cos}(\zeta)
    \end{pmatrix},
\end{align}

\begin{eqnarray}
    J(\zeta)=\begin{pmatrix}
       \displaystyle  \frac{2}{b_r}p_x \\ \displaystyle \frac{2}{b_r}p_y \\ \displaystyle \frac{2p_z}{b_z}-\frac{d}{b_z^2} 
    \end{pmatrix} + \displaystyle D^T(\zeta)\begin{pmatrix}
      \displaystyle  \frac{2}{b_r}q_x \\ \displaystyle \frac{2}{b_r}q_y \\\displaystyle  \frac{2q_z}{b_z}-\frac{d'}{b_z^2} 
    \end{pmatrix}, 
\end{eqnarray}  
\begin{eqnarray}    
    \displaystyle M(\zeta) = B^2+D^T(\zeta)B^2D(\zeta).
\end{eqnarray}

\noindent The explicit expressions of $J(\zeta)$ and $M(\zeta)$ are given in Appendix \ref{appendixa}.
With these definitions, the spatial matrix element takes the form:

\begin{equation}
\begin{array}{lcl}
    R_{\nu\nu'}(\zeta) &=& \displaystyle  \mathcal{N}_{n_z}\mathcal{N}_{n_-n_+}\mathcal{N}_{n_z'}\mathcal{N}_{n_+'n_-'} \\ \displaystyle & \times& \displaystyle  \frac{\displaystyle \partial^{n_z}\partial^{n_-}\partial^{n_+}}{\partial p_z^{n_z}\partial p_+^{n_-}\partial p_-^{n_+}}  \frac{\displaystyle \partial^{n'_z}\partial^{n'_+}\partial^{n'_-}}{\partial q_z^{n'_z}\partial q_+^{n'_+}\partial q_-^{n'_-}} \nonumber \\ &\times& \displaystyle e^{\displaystyle \left[-p^2 - q^2+\frac{2p_zd+2q_zd'}{b_z}\right]}    \nonumber  \\
 &\times& \displaystyle  \left. \int \; d\vec{r} \;  e^{\displaystyle \left[-\frac{1}{2}\vec{r}M(\zeta)\vec{r}+J(\zeta)^T \vec{r}\right]} \right|_{p=q=0}. \nonumber
\end{array}
\end{equation}

\noindent Since the matrix $M(\zeta)$ is symmetric and positive definite, the Gaussian integral can be evaluated analytically. We obtain:
\begin{align}
    \int \; d\vec{r} \;  e^{\displaystyle \left[-\frac{1}{2}\vec{r}M(\zeta)\vec{r}+J(\zeta)^T \vec{r}\right]} = \sqrt{\frac{8\pi^3}{\text{det}(M(\zeta))}}  \\ \displaystyle 
    \times e^{\displaystyle \left[\frac{1}{2}J(\zeta)^TM(\zeta)^{-1}J(\zeta)\right]}. \nonumber
\end{align}

\noindent The resulting expression for $R_{\nu\nu'}(\zeta)$ can therefore be written as:

\begin{widetext}
\begin{eqnarray}
    & \displaystyle R_{\nu\nu'}(\zeta) = \displaystyle \mathcal{N}_{n_z}\mathcal{N}_{n_-n_+}\mathcal{N}_{n_z'}\mathcal{N}_{n_+'n_-'}\sqrt{\frac{8\pi^3}{\text{det}(M(\zeta))}} \left.\frac{\partial^{n_z}\partial^{n_-}\partial^{n_+}}{\partial p_z^{n_z}\partial p_+^{n_-}\partial p_-^{n_+}}  \frac{\partial^{n'_z}\partial^{n'_+}\partial^{n'_-}}{\partial q_z^{n'_z}\partial q_+^{n'_+}\partial q_-^{n'_-}}  I_G(\zeta)\right|_{p=q=0}
\end{eqnarray}
\noindent where the integral $I_G(\zeta)$ is defined by:
\begin{eqnarray}
    &I_G(\zeta) = e^{\displaystyle \left[ -p^2 - q^2+\frac{2p_zd+2q_zd'}{b_z}+  \frac{1}{2}J(\zeta)^TM(\zeta)^{-1}J(\zeta) \right]}.
\end{eqnarray}
\end{widetext}

The extraction of the spatial matrix elements requires the successive evaluation of several partial derivatives of $I_G(\zeta)$. A brute-force evaluation would lead to a 
combinatorial proliferation of terms, making the resulting expressions increasingly cumbersome to handle analytically and severely hampering, if not preventing, their numerical evaluation. 
To circumvent this issue, we introduce the multivariate Hermite polynomials which allow a recursive evaluation of the required partial derivatives. This approach substantially simplifies the analytical derivation while significantly improving the numerical efficiency of the calculation at the same time. \\

Defining the quantity $X^T = \left( p_+ ~ p_- ~ p_z ~ q_+ ~ q_- ~ q_z \right) $,
the integral $I_G(\zeta)$ can then be factorized as:
\begin{align}
    I_G(\zeta)= e^{\displaystyle \left[-X^TA(\zeta)X + B(\zeta)^TX + C(\zeta)\right]}, 
\end{align}
\noindent The explicit expressions of the matrix $A(\zeta)$, the vector $\; B(\zeta)$ and the scalar $C(\zeta)$ are given in Appendix \ref{appendixa}.

At this point, we introduce the multivariate Hermite polynomials $H_{(\alpha_i,...,\alpha_6)}(A(\zeta),B(\zeta),X)$ whose expression is:

\begin{widetext}
\begin{eqnarray}
    H_{(\alpha_1,...,\alpha_6)}\left(A(\zeta),B(\zeta),X\right) = (-1)^{\alpha_1+...+\alpha_6}    ~ e^{\displaystyle \left[X^TA(\zeta)X - B(\zeta)^TX\right]} \qquad \qquad \qquad \qquad \qquad \qquad \qquad \qquad \\ 
\times \frac{\displaystyle \partial^{\alpha_1}\partial^{\alpha_2}\partial^{\alpha_3}}{\partial p_+^{\alpha_1}\partial p_-^{\alpha_2}\partial p_z^{\alpha_3}} \frac{\displaystyle \partial^{\alpha_4}\partial^{\alpha_5}\partial^{\alpha_6}}{\partial q_+^{\alpha_4}\partial q_-^{\alpha_5}\partial q_z^{\alpha_6}} ~ e^{\displaystyle \left[-X^TA(\zeta)X+B(\zeta)^TX\right]}.\nonumber
\end{eqnarray}

\noindent These polynomials satisfy, in particular, the following recurrence relation:
\begin{eqnarray}\label{MVHP}
   \displaystyle  H_{(\alpha_1,...,\alpha_j+1,...,\alpha_6)}\left(A(\zeta),B(\zeta),0\right) =  B(\zeta)_j ~ H_{(\alpha_1,...,\alpha_6)}(A(\zeta),B(\zeta),0) \qquad \qquad \qquad \qquad \\ - 2\sum_l\alpha_l A(\zeta)_{lj} ~H_{(\alpha_1,...,\alpha_l-1,...,\alpha_6)}\left(A(\zeta),B(\zeta),0\right). \nonumber
\end{eqnarray}

\noindent Using Eq.\eqref{MVHP}, the expression of the matrix elements $R_{\nu\nu'}(\zeta)$ finally transforms into:
\begin{eqnarray}
    R_{\nu\nu'}(\zeta) = \mathcal{N}_{n_z}\mathcal{N}_{n_-n_+}\mathcal{N}_{n_z'}\mathcal{N}_{n_+'n_-'}~ e^{\displaystyle \left[C(\zeta)\right]} \sqrt{\displaystyle \frac{8\pi^3}{\text{det}(M(\zeta))}}  (-1)^{n_z+n_-+n_++n_z'+n'_++n_-'}  \qquad \qquad \qquad \\ \times H_{(n_-,n_+,n_z,n'_+,n'_-,n'_z)}(A(\zeta),B(\zeta),0). \nonumber
\end{eqnarray}
\end{widetext}

\subsubsection{Spin contribution}

The spin contribution to $R_{\nu\nu'}(\zeta)$ is easily deduced. We consider the action of a rotation by an angle $\zeta$ about the $y$-axis on the spin eigenstates $| \pm \rangle $ of the operator $\hat s_z$:
\begin{eqnarray}
    &e^{\displaystyle -i\zeta \hat s_y} |+\rangle = \text{cos}(\frac{\zeta}{2}) ~|+\rangle + \text{sin}(\frac{\zeta}{2}) ~|-\rangle, \nonumber \\  &e^{\displaystyle -i\zeta \hat s_y} ~|-\rangle = \text{cos}(\frac{\zeta}{2})|-\rangle - \text{sin}(\frac{\zeta}{2}) ~|+\rangle.
\end{eqnarray}

\noindent The previous relations directly provide the intrinsic spin contribution $S_{s_\nu s_{\nu'}}(\zeta)$ to the matrix elements of the rotation operator:
\begin{eqnarray}
    S_{++}(\zeta) &=& + \text{cos}\left(\frac{\zeta}{2}\right), \\ S_{--}(\zeta) &=& + \text{cos}\left(\frac{\zeta}{2}\right),\\
    S_{-+}(\zeta) &=& + \text{sin}\left(\frac{\zeta}{2}\right), \\S_{+-}(\zeta) &=& -\text{sin}\left(\frac{\zeta}{2}\right).  
\end{eqnarray}

\subsubsection{Symmetries}

In this section, we enumerate the symmetries of $R^s_{\nu\nu'}(\zeta)$ that will be used in the rest of the derivations. We first remark that the reality of $R_{\nu\nu'}(\zeta)$ implies that:
    \begin{eqnarray}
      R_{(m,n,n_z,d)(m',n',n_z',d')}(\zeta) = R_{(-m,n,n_z,d)(-m',n',n_z',d')}(\zeta) 
    \end{eqnarray}
    and
    \begin{eqnarray}
        R_{\nu \nu'}(\zeta) = R_{\nu' \nu}(-\zeta).
    \end{eqnarray}

\noindent Then, noting that the change of variable $x'=-x$ is the same as replacing $m$ by $-m$, we obtain:
    \begin{eqnarray}
    R_{\nu \nu'}(\zeta) &=& (-1)^{n_+ + n_- + n'_+ + n'_-} R_{\nu'\nu}(\zeta), \\  & = & (-1)^{n_+ + n_- + n'_+ + n'_-} R_{\nu\nu'}(-\zeta). \nonumber  
    \end{eqnarray}
    
Finally, the parity under the transformation $z\rightarrow-z$ provides:
    \begin{eqnarray}
        R_{\left(\nu_d,\nu'_{d'}\right)}(\zeta) = (-1)^{m+n_z+m'+n_z'}R_{\left(\nu_{-d},\nu'_{-d'}\right)}(\zeta).
    \end{eqnarray}

Choosing the following convention for the time-reversal phases, for $m_\nu+s_\nu>0$,
\begin{eqnarray}
    \displaystyle  |\bar \nu \rangle &=& \displaystyle \sigma_\nu|-m_\nu,n_\nu,n_{z_\nu},\bar s_\nu \rangle, \\  \sigma_\nu &=&(-1)^{\frac{\displaystyle 1}{\displaystyle  2} \displaystyle  +\bar s_\nu},
\end{eqnarray}
\noindent we obtain the two following relations:
\begin{eqnarray}
    R^s_{\nu \nu'}(\zeta) &=& R^s_{\bar \nu \bar \nu'}(\zeta) \label{eq43}
\end{eqnarray}
\noindent and
\begin{eqnarray}
    R^s_{\bar \nu \nu'}(\zeta) &=& - R^{s}_{\nu \bar \nu'}(\zeta) \label{eq44}.
\end{eqnarray}

\noindent Equations \eqref{eq43} and \eqref{eq44} lead to the following block structure for the matrix $R^s$:
\begin{eqnarray}
    &R^s(\zeta) = \begin{pmatrix}
        R^s_{\Omega \Omega}(\zeta) & R^s_{\Omega \bar \Omega}(\zeta) \\
        -R^s_{\Omega \bar \Omega}(\zeta) & R^s_{\Omega \Omega}(\zeta)
    \end{pmatrix}
\end{eqnarray}
where $\Omega$ denotes the subspace associated with $m_\nu+s_\nu >0$ and $\bar \Omega$ the one satisfying $m_\nu+s_\nu <0$.

\subsection{Evaluation of the rotated overlap kernel}\label{sub2ROTA}

The aim of this section is to derive an analytical expression for the rotated overlap kernel $\langle \Phi|\Phi(\zeta)\rangle$. Building on the established pfaffian framework for the evaluation of such overlaps \cite{RobPfaff,RobCB,BerRobPfaff,Rob2025}, we develop a new formulation incorporating two important simplifications. Firstly, it avoids enlarging the single-particle space to accommodate the rotated basis. Secondly, we show that, owing to the symmetries of the problem, the pfaffian can be reduced to the determinant of an $n\times n$ complex matrix. This provides a convenient numerical implementation based on standard, highly optimized linear-algebra routines and offers improved numerical efficiency and robustness. \\

We first recall that, under axial symmetry and time-reversal invariance, the Hartree-Fock-Bogoliubov transformation can be written as:
\begin{eqnarray}
    \xi_i &=& \sum_{k} U_{ki}c_k + V_{ki}\bar c_k^+, \\ 
    \bar \xi_i &=& \sum_{k} U_{ki}\bar c_k - V_{ki}c_k^+,
\end{eqnarray}
where the set of $\{\xi_i\}$ represent the quasiparticle annihilation operators. Using the standard commutator expansion together with the symmetries of the matrix $R^s(\zeta)$, we obtain:
\begin{eqnarray}
    e^{-i\zeta \hat J_y} ~c_k~ e^{i\zeta \hat J_y} &=& \sum_{\nu} R^s_{\nu' k}(\zeta) ~c_{\nu'}, \\
    e^{-i\zeta \hat J_y} ~c^+_k~ e^{i\zeta \hat J_y} &=& \sum_{\nu} R^s_{\nu' k}(\zeta) ~c^+_{\nu'}.
\end{eqnarray}
\noindent We then express the rotated norm kernel $\langle \Phi|\Phi(\zeta)\rangle$ using the Thouless theorem, assuming that the matrix $U$ is invertible:
\begin{align}
     \langle \Phi| \Phi(\zeta)\rangle = \left|\langle\Phi| 0\rangle \right|^2 \langle 0| ~e^{\displaystyle \frac{1}{2}\sum_{kk'}c_kc_{k'}(\tilde V \tilde U^{-1})_{kk'}} \\ \nonumber \times e^{\displaystyle -i\zeta\hat J_{y}} ~e^{\displaystyle \frac{1}{2}\sum_{kk'}(\tilde V \tilde U^{-1})_{kk'}c^+_kc^+_{k'}} |0 \rangle,
\end{align}
where $\tilde U$ and $\tilde V$ are the corresponding full Bogoliubov transformation matrices:
\begin{align}
    \tilde U = \begin{pmatrix}
        U & 0 \\ 0 & U
    \end{pmatrix}, && \tilde V = \begin{pmatrix}
        0 & -V \\ V & 0
    \end{pmatrix}.
\end{align}

\noindent As shown in \cite{TPaul,trilogy1}, this expression can be re-casted as:
\begin{eqnarray}\label{equation54}
     \langle \Phi| \Phi(\zeta)\rangle =|\langle\Phi| 0\rangle|^2 \langle 0|e^{\displaystyle \frac{1}{2}\sum_{kk'}c_kc_{k'}M^{(1)}_{kk'}} \qquad \qquad \qquad \\ \nonumber \times  e^{\displaystyle \frac{1}{2}\sum_{kk'}\big(M^{(0)}(\zeta)\big)_{kk'}c^+_kc^+_{k'}} |0 \rangle,
\end{eqnarray}
\noindent with 
\begin{eqnarray}
    \displaystyle M^{(0)}(\zeta) &=& \displaystyle R^s(\zeta)\tilde V \tilde U^{-1}R^{sT}(\zeta), \\ \displaystyle M^{(1)} &=& \displaystyle \tilde V \tilde U^{-1}.
\end{eqnarray}
Although the rotation formally requires the sums in the exponential on the right-hand side of Eq.\eqref{equation54} to run over the complete single-particle space, thereby extending the basis associated with the $\{c_k\}$ operators, the terms involving these additional indices vanish as a consequence of the anti-commutation relations satisfied by the single-particle operators \cite{TPaul,trilogy1}. 
With the help of Grassmann algebra, \cite{RobPfaff,TPaul,trilogy1}, this expression can be reduced to:
\begin{eqnarray}
    \displaystyle \langle\Phi|\Phi(\zeta)\rangle &=& (-1)^{\frac{\displaystyle n(n+1)}{\displaystyle 2}}  ~ \left|\langle \Phi|0\rangle \right| \\ && \qquad \qquad \times \text{pf}\big(M^{(1)}\big) \text{pf}\big(M(\zeta)\big), \nonumber
\end{eqnarray}
where $pf$ represents the pfaffian and the matrix $M(\zeta)$ is defined by:
\begin{align}
    M(\zeta) = M^{(0)}(\zeta)+M^{(1)-1}.
\end{align}
\noindent The block structure of the matrix $M^{(1)}$ allows us to express its pfaffian in the following way:
\begin{align}
    &\text{pf}\big(M^{(1)}\big) = (-1)^{\frac{\displaystyle n}{\displaystyle 4}(\frac{\displaystyle n}{\displaystyle 2}\displaystyle -1)}(-1)^{\frac{\displaystyle n}{\displaystyle 2}} ~ \text{det}\big(VU^{-1}\big).
\end{align}
We also use the well-known result \cite{rsbook,TPaul}:
\begin{eqnarray}
    \left|\langle\Phi|0\rangle \right| &=& \text{det}\big(U\big)^2.
\end{eqnarray}
Furthermore, the matrix $M(\zeta)$ has the particular block structure:
\begin{align}
    M(\zeta) = \begin{pmatrix}
       X(\zeta)  & Y(\zeta) \\ -Y(\zeta) & X(\zeta)
        \end{pmatrix},
\end{align}
where the antisymmetric sub-matrix $X(\zeta)$ is given by:
\begin{eqnarray}
\displaystyle X(\zeta) = R^s_{\Omega \Omega}(\zeta)VU^{-1}R^{sT}_{\Omega \bar \Omega}(\zeta) \qquad \qquad \qquad \\ \displaystyle  - R^s_{\Omega \bar \Omega}(\zeta)VU^{-1}R^{sT}_{\Omega  \Omega}(\zeta),  \nonumber 
\end{eqnarray}
while the symmetric sub-matrix $Y(\zeta)$ has for expression:
\begin{eqnarray}
    Y(\zeta) = R^s_{\Omega\Omega}(\zeta)VU^{-1}R^{sT}_{\Omega\Omega}(\zeta) \qquad \qquad \qquad \\ \nonumber + R^s_{\Omega\bar \Omega}(\zeta)VU^{-1}R^{sT}_{\Omega\bar \Omega}(\zeta) + UV^{-1}.
\end{eqnarray}
\noindent This structure allows $M(\zeta)$ to be factorized as:
\begin{align}
     &M(\zeta)  = T^* \begin{pmatrix}
        0 & X(\zeta) + iY(\zeta) \\ - (X(\zeta)+iY(\zeta))^T & 0
    \end{pmatrix} T^{\dagger},
\end{align}
where $T^{\dagger}$ is the Hermitian conjugate of $T$:
\begin{align}
 \displaystyle T= \frac{\displaystyle 1}{\displaystyle \sqrt{2}}\begin{pmatrix}
       \displaystyle  I & \displaystyle I \\ \displaystyle -iI &\displaystyle  iI
    \end{pmatrix}.
\end{align}
Then, this factorization yields directly: 
\begin{eqnarray}
    \displaystyle \text{pf}\big(M(\zeta)\big) =  (-1)^{\frac{\displaystyle n}{\displaystyle 4}(\frac{\displaystyle n}{\displaystyle 2}\displaystyle -1)}(-i)^{\frac{\displaystyle n}{\displaystyle 2}} \qquad \qquad \\ \times ~ \text{det}\big(X(\zeta)+iY(\zeta)\big). \nonumber
\end{eqnarray}
This finally leads to the following expression for $\langle\Phi|\Phi(\zeta)\rangle$:
\begin{eqnarray}
    \displaystyle \langle\Phi|\Phi(\zeta)\rangle &=& (-i)^{\frac{\displaystyle n}{\displaystyle 2}} ~ \text{det}\big(V^T(X(\zeta)  + i Y(\zeta))U\big).~~~~
\end{eqnarray}

\subsection{Gauss-Legendre Quadrature}\label{sub3QUAD}

The total AM projected norm can now be evaluated using Gauss-Legendre quadrature. Starting from Eq.\eqref{eqPJ},
we make the change of variables:
\begin{eqnarray}
x &=& \text{cos}(\zeta), \\ dx &=& - \text{sin}(\zeta) ~ d\zeta.
\end{eqnarray}
\noindent This gives:
\begin{eqnarray}
    \langle \Phi | \hat P^{J} | \Phi\rangle = \frac{2J + 1}{2}  \int^1_{-1}  e^{\displaystyle \left[\text{arccos}(x)i J \right]} \qquad \qquad
    \\ \nonumber \times P_J(x)\langle \Phi|\Phi\left(\text{arccos}(x)\right)\rangle ~dx.
\end{eqnarray}
Then, applying a Gauss-Legendre quadrature, one obtains:
\begin{eqnarray}
    \displaystyle \langle \Phi|\hat P^{J}|\Phi\rangle \approx \frac{2J + 1}{2}  \sum_{k=1}^{n_{GL}} w_{GL_k} e^{\displaystyle \left[\text{arccos}(x_k)i J\right]}  \qquad \qquad \\ \nonumber  \times P_J(x_k)\langle \Phi|\Phi\left(\text{arccos}(x_k)\right)\rangle.
\end{eqnarray}

\subsection{Inclusion of particle number projection}\label{sub4PN}

When particle-number projection is included in addition to AM projection, the relevant overlap becomes
\begin{align}
     \langle \Phi| \Phi(\zeta,\varphi)\rangle = \left|\langle\Phi| 0\rangle \right|^2 \langle 0| ~e^{\displaystyle \frac{1}{2}\sum_{kk'}c_kc_{k'}(\tilde V \tilde U^{-1})_{kk'}} \nonumber \\ \times e^{\displaystyle -i\zeta\hat J_{y}}  ~e^{\displaystyle -i\varphi\hat N}  ~e^{\displaystyle \frac{1}{2}\sum_{kk'}(\tilde V \tilde U^{-1})_{kk'}c^+_kc^+_{k'}} |0 \rangle,
\end{align}
where $\hat N$ denotes the particle-number operator. For simplicity, the isospin degree of freedom is omitted here. Following the same procedure as in section \ref{sub2ROTA}, the overlap can be recast as
\begin{eqnarray}
     \langle \Phi| \Phi(\zeta,\varphi)\rangle =|\langle\Phi| 0\rangle|^2 \langle 0|e^{\displaystyle \frac{1}{2}\sum_{kk'}c_kc_{k'}M^{(1)}_{kk'}} \qquad \qquad \qquad \\ \nonumber \times  e^{\displaystyle \frac{1}{2}\sum_{kk'}\big(M^{(0)}(\zeta,\varphi)\big)^*_{kk'}c^+_kc^+_{k'}} |0 \rangle,
\end{eqnarray}
\noindent with
\begin{eqnarray}
    \displaystyle M^{(0)}(\zeta,\varphi) &=& \displaystyle e^{2i\varphi}M^{(0)}(\zeta).
\end{eqnarray}
Introducing the symmetric matrix $Y(\zeta,\varphi)$ 
\begin{eqnarray}
    Y(\zeta,\varphi) = R^s_{\Omega\Omega}(\zeta)VU^{-1}R^{sT}_{\Omega\Omega}(\zeta) \qquad \qquad \qquad \\ \nonumber + R^s_{\Omega\bar \Omega}(\zeta)VU^{-1}R^{sT}_{\Omega\bar \Omega}(\zeta) + e^{2i\varphi}UV^{-1},
\end{eqnarray}
and following the same procedure as in section \ref{sub2ROTA}, the final expression for the overlap is
\begin{align}
     \langle \Phi| \Phi(\zeta,\varphi)\rangle = (-1)^{\frac{n}{2}}e^{-in\varphi} \det\big((V^T(X(\zeta)+ iY(\zeta,\varphi))U \big).
\end{align}

Once this overlap has been evaluated, the only modification to the previous method concerns the numerical integration over the gauge angle $\varphi$, which is performed using a uniform discretization of the integration interval.

\section{Total angular momentum distribution of fission fragments in $^{240}$Pu }\label{section4}

In this section, the total AM projection formalism developed in section \ref{theory} is applied to an asymmetric adiabatic fission path of $^{240}$Pu. The main objective is to characterize the total AM distributions of the fragments in the vicinity of scission. The analysis combines total AM projection with particle number projection, thereby allowing the total AM distributions to be resolved according to the different fragmentations contained in the HFB states. The one-dimensional fission path considered here is obtained using the two recently proposed \enquote{Link} and \enquote{Drop} methods \cite{carpentier2024,trilogy1}. These methods are based on HFB calculations with overlap constraints and are designed to provide a continuous and regular description of the fission path through the scission region and beyond. The resulting path therefore provides a suitable framework for following the evolution of the fragment total AM distributions across the scission region and for examining their dependence on the underlying HFB configuration.

The application of total AM projection to the fission fragments first requires a consistent definition of the two fragments within the HFB state. This point is particularly important in the present framework, since the fragment total AM operators must remain compatible with the single-particle basis in which the matrix elements of the rotation operator are calculated. The first part of this section therefore introduces the fragment separation procedure adopted here, which is based on the canonical basis separation method developed in \cite{TPaul,trilogy1} and inspired by the quantum separation proposed in \cite{Rota}. This construction provides a consistent framework in which particle number and AM projection can be performed independently for the two fragments.

The resulting total AM distributions are then analyzed for both the light and heavy fragments. Particular attention is devoted to the evolution of the distributions in the scission region and to the emergence of a pronounced odd-even structure, characterized by an increasing dominance of even-$J$ components as the system approaches scission. Its relation to the intrinsic shape of the fragments is investigated through the evolution of their multipole moments, in particular the intrinsic octupole moment $Q_{30,F}^{CM}$. Finally, particle-number projection is used to resolve the AM distributions according to the different fragmentations present in the HFB states. This decomposition makes it possible to investigate the role of structure effects, in particular shell effects associated with specific fragment configurations, in shaping the total AM distributions.

\subsection{A consistent fragment separation}

The fragment separation at scission used in this work is based on the method described in \cite{TPaul,trilogy1}. This method consists in separating the quasiparticle operators in the canonical basis $\{\eta\}$ into two families, $\{\eta_l\}_{l\in L}$ and $\{\eta_h\}_{h\in H}$, associated with the light and heavy fragments, respectively. The HFB state can then be written as a tensor product,
\begin{align}
    |\Phi\rangle = |\Phi_L\rangle \otimes | \Phi_H\rangle.
\end{align}
Since the separation is performed in the canonical basis, the corresponding particle operators $\{a,a^+\}$ are also separated into two families, $\{a_l,a_l^+\}_{l\in L}$ and $\{a_h,a_h^+\}_{h\in H}$. The particle number operators associated with the light $\hat N_L$ and heavy $\hat N_H$ fragments therefore take the simple form:
\begin{equation}
\begin{array}{lcl}
   \hat N_L &=& \displaystyle \sum_{l\in L}a_l^+a_l, \\
   \hat N_H &=& \displaystyle \sum_{h\in H}a_h^+a_h. 
\end{array}
\end{equation}
\noindent Similarly, the total AM operators of the light and heavy fragments can be written as:
\begin{equation}
\begin{array}{lcl}
    \hat J_L &=& \displaystyle \sum_{ll'\in L}\langle l| \hat j_L|l'\rangle a^+_la_{l'} , \\ 
    \hat J_H &=& \displaystyle \sum_{hh'\in H}\langle h| \hat j_H|h'\rangle a^+_ha_{h'}.
\end{array}
\end{equation}
Here, $\hat j_L$ and $\hat j_H$ denote the corresponding single-particle total AM operators defined with respect to the centers of mass of the light and heavy fragments, respectively. In the calculations presented below, which combine particle number and AM projection, the fragment centers of mass are those obtained after particle number projection.

This separation offers two main advantages over the separation schemes commonly used in the literature, in particular the one recently employed in \cite{marevic2025,marevic2026}. The first one is of a practical and numerical nature. Because the state factorizes into a product of two HFB states, the light and heavy fragment operators act in two separate Hilbert spaces. The corresponding particle number and AM projectors can therefore be applied independently to each fragment. In particular,
\begin{eqnarray}
    \hat P^{J_L}\hat P^{J_H} \hat P^{N_L}\hat P^{N_H} |\Phi \rangle =  \hat P^{J_L} \hat P^{N_L}|\Phi_L\rangle \qquad \qquad \qquad \\ \otimes \hat P^{J_H}\hat P^{N_H}| \Phi_H\rangle. \nonumber
\end{eqnarray}

\noindent The second advantage is more fundamental. The single-particle basis in which the fragment AM operators $\hat J_L$ and $\hat J_H$ are defined is in one-to-one correspondence with the 2-center basis used to evaluate the matrix elements of the single-particle rotation operator. This feature avoids the inconsistencies identified by L.M. Robledo in \cite{robledo2026} of the separation schemes previously employed in the literature. 

The proposed separation method also has some limitations. In particular, it cannot describe odd fragmentations, as the construction imposes a factorization into HFB quasiparticle vacua and therefore restricts each fragment to an even particle-number parity. Furthermore, the separation does not result in a strictly spatial partition of the system. Some particles assigned to the light fragment remain localized in the spatial region of the heavy one, and vice versa. These effects are quantified and discussed in \cite{TPaul,trilogy1}. 

Finally, the scission region in which the calculations are performed is determined as described in \cite{TPaul,trilogy1}, from the peak of the chemical potential. In the present case, this criterion places the scission region around $Q_{20}=13110~\mathrm{fm}^2$.

\subsection{Results and Analysis}

\begin{figure}
\centering
\includegraphics[width=1.0\linewidth]{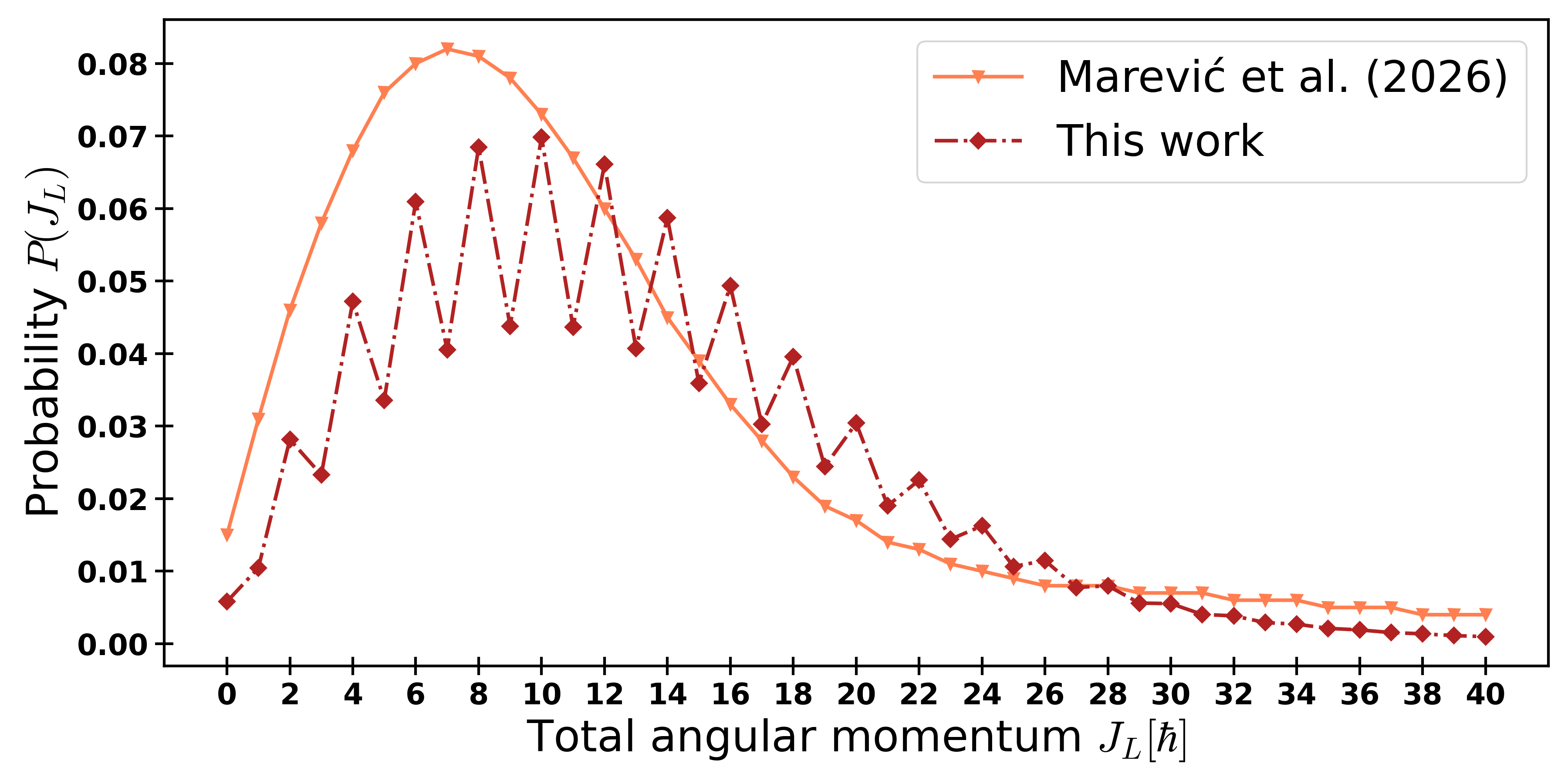}
\caption{Total angular momentum distribution of the light fragment at $Q_{20} = 13110 \;fm^2$, obtained with particle-number projection and averaged over the different fragmentations.}
\label{figMareL}
\end{figure}

\begin{figure}
\centering
\includegraphics[width=1.0\linewidth]{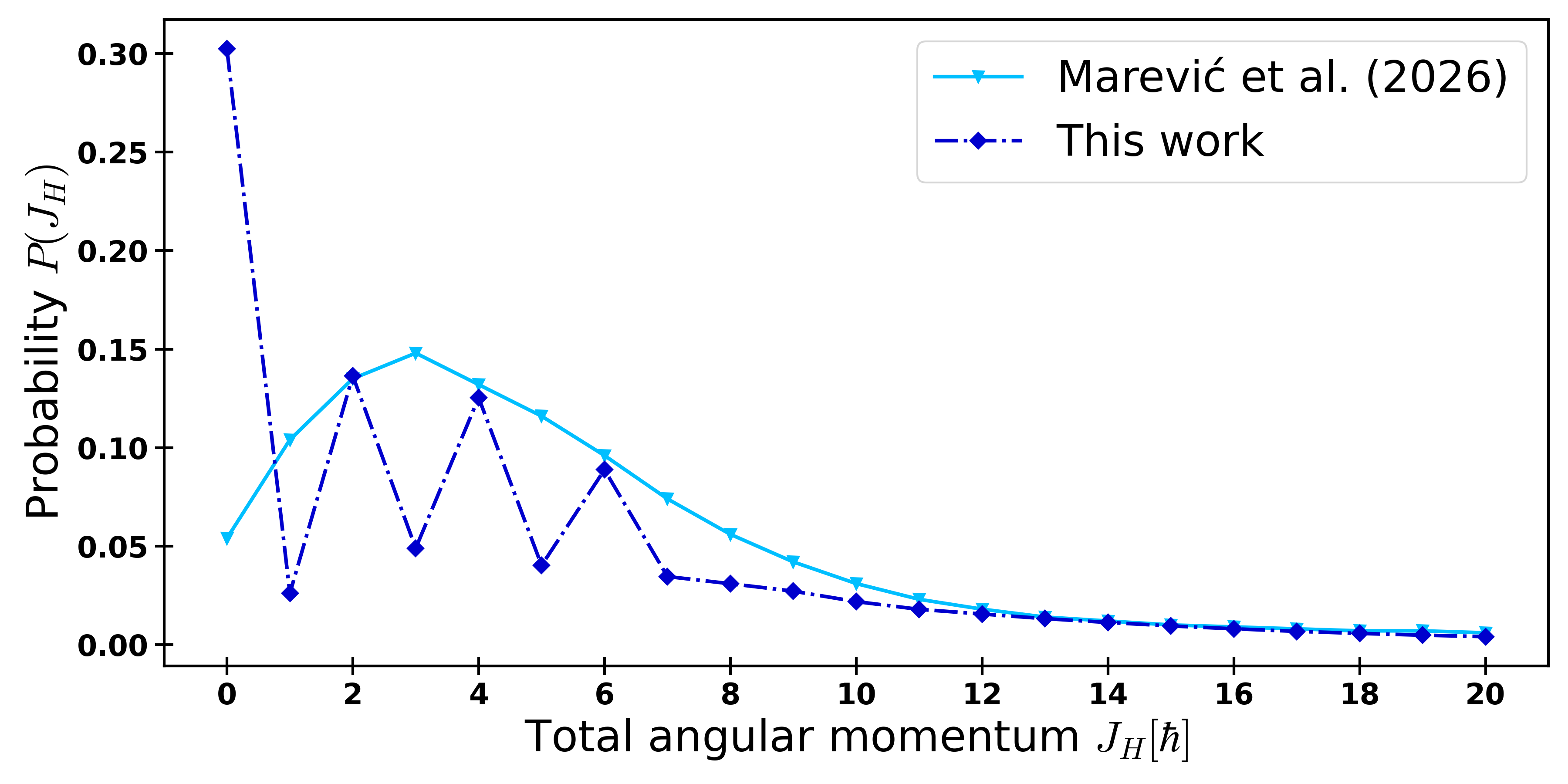}
\caption{Same as FIG. \ref{figMareL}, but for the heavy fragment.}
\label{figMareR}
\end{figure}

To assess the robustness of the present approach, the total AM distributions of the fragments at $Q_{20}=13110~\mathrm{fm}^2$, obtained with the particle number projection and averaged over the different fragmentations, are first compared with the recent results of P. Marević et \textit{al.} \cite{marevic2025} in FIGs. \ref{figMareL} and \ref{figMareR}. This study provides one of the closest available references to the present calculation, as it considers an adiabatic HFB state associated with the most probable exit point of a TDGCM+GOA calculation. In this respect, its physical setting is conceptually close to the HFB asymmetric minimum-energy path considered here.


Several differences between the two calculations should nevertheless be kept in mind when interpreting the comparison. In addition to the different prescriptions used to separate the two fragments, the average fragmentation considered here, $(N_H=82,Z_H=51.5)$, differs from the one of Marević et \textit{al.}, $(N_H=84.8,Z_H=53.5)$. Moreover, the configuration considered by Marević et \textit{al.} corresponds to a substantially larger neck density, $Q_{\rm neck}\approx 2.5$, whereas the configuration associated with the peak of the chemical potential along the present path has $Q_{\rm neck}\approx 0.3$. These differences preclude a detailed quantitative comparison of the individual distributions. The comparison is therefore intended primarily as a qualitative assessment of the consistency of the two approaches.

Despite these differences, the total AM distributions obtained in the two calculations exhibit similar overall features. In particular, both approaches predict a pronounced asymmetry in the sharing of AM between the two fragments, with the light fragment carrying the larger total AM. The average angular momenta obtained here, $\langle J_L\rangle=12.98\hbar$ and $\langle J_H\rangle=4.5\hbar$, are also of the same order as those reported by Marević et \textit{al.}, $\langle J_L\rangle=10.4\hbar$ and $\langle J_H\rangle=5.4\hbar$. This overall agreement provides a useful indication that the present projection and fragment separation procedure captures the main features of the fragment total AM sharing. \\

\begin{figure}
\centering
\includegraphics[width=1.0\linewidth]{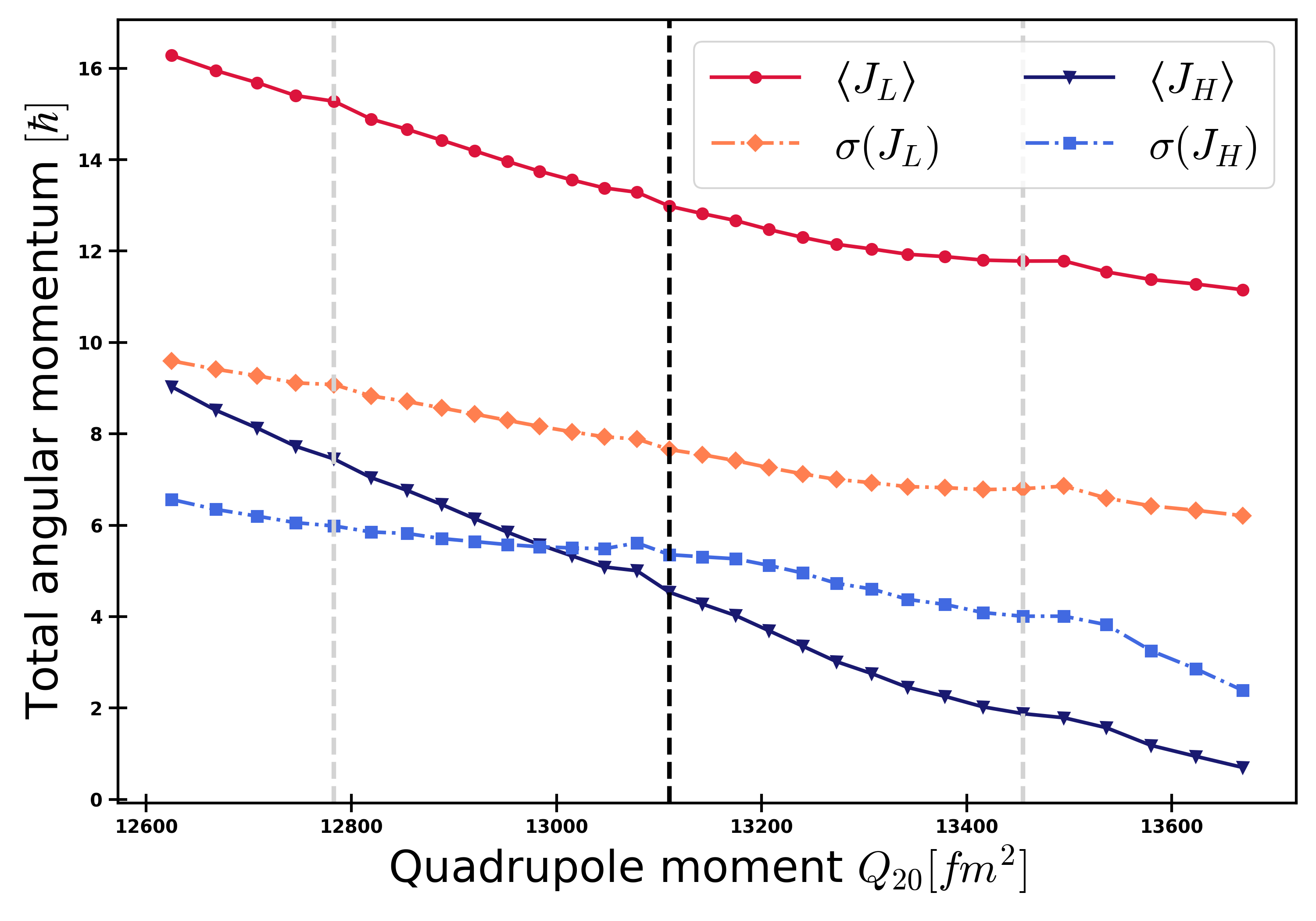}
\caption{Evolution of $\langle J_L\rangle$, $\langle J_H\rangle$, $\sigma(J_L)$, and $\sigma(J_H)$ as a function of $Q_{20}$.}
\label{figIndic}
\end{figure}

\begin{figure}
\centering
\includegraphics[width=0.8\linewidth]{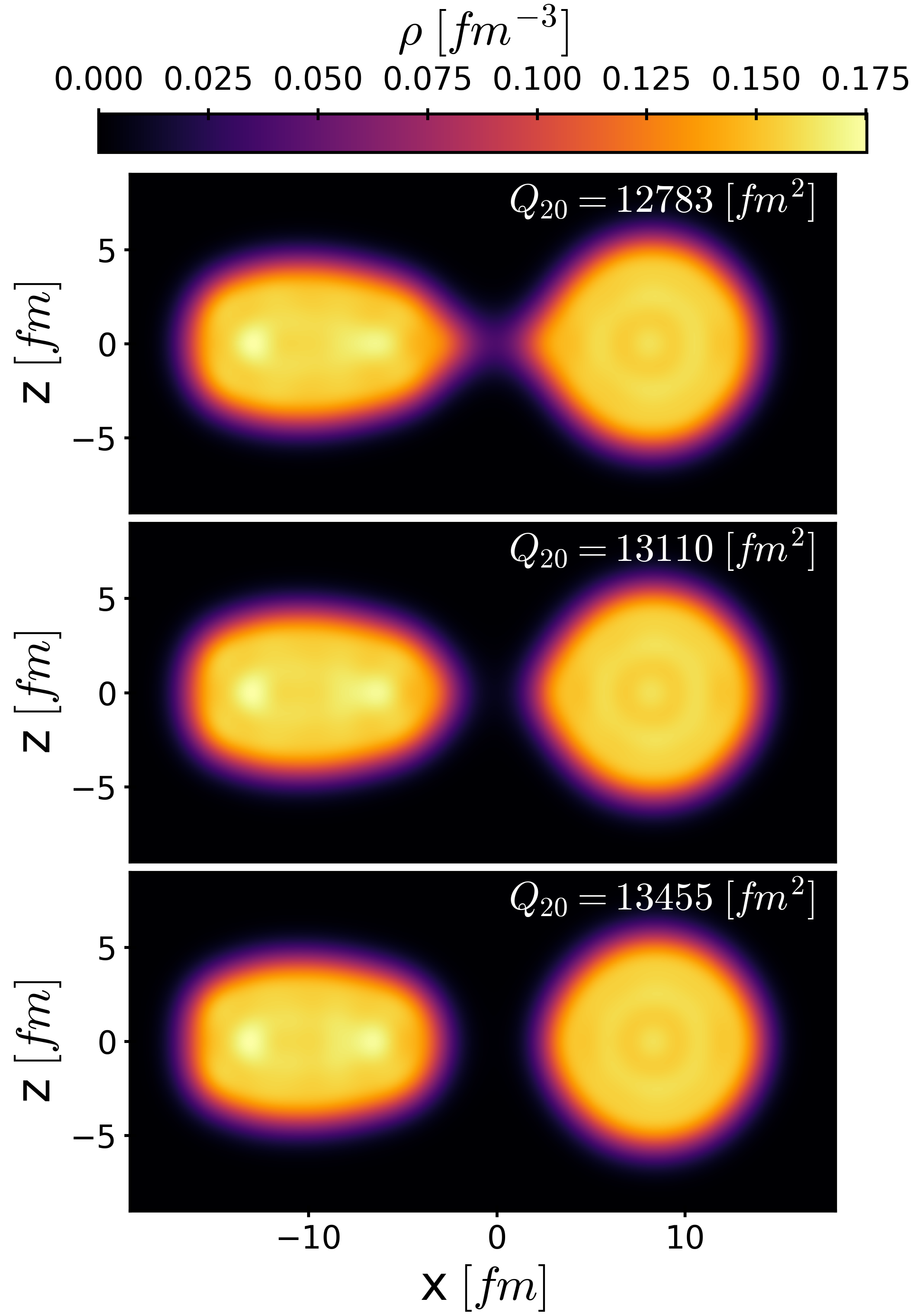}
\caption{Local densities of the configurations characterized by $Q_{20}=12783~\mathrm{fm}^2$, $Q_{20}=13110~\mathrm{fm}^2$, and $Q_{20}=13455~\mathrm{fm}^2$.}
\label{figDen}
\end{figure}

Beyond this overall agreement, one may now investigate the origin of the remaining differences between the distributions, including the differences in their statistical moments, the pronounced dominance of even-$J$ components observed in the present results but not in those of Marević et \textit{al.}, and the large value of $P(J_H=0)$. This naturally raises the question of the dependence of the total AM distributions on the choice of the HFB state at which they are evaluated. To obtain a first qualitative estimate of the effect of the scission point choice, FIG. \ref{figIndic} shows the evolution of $\langle J_L\rangle$, $\langle J_H\rangle$, $\sigma(J_L)$, and $\sigma(J_H)$ as a function of quadrupole deformation $Q_{20}$ of the fissioning nucleus.
The vertical black line indicates the configuration at the peak of the chemical potential, which is the configuration used for the results shown in FIGSs. \ref{figMareL} and \ref{figMareR}. The two gray vertical lines on each side correspond to configurations characterized by $Q_{\rm neck}\approx0.7$ and $Q_{\rm neck}\approx0.1$. The corresponding local densities are shown in FIG. \ref{figDen} and are provided as a qualitative reference to assess the evolution of the global shape of the nucleus in the scission region.

A pronounced, approximately linear dependence of the statistical properties of the total AM distributions on $Q_{20}$ is observed. This behavior emphasizes that the total AM distributions of the fragments can depend significantly on the HFB configuration at which they are evaluated. Since both $\langle J_L\rangle$ and $\langle J_H\rangle$ decrease with increasing $Q_{20}$, and therefore implicitly with decreasing $Q_{\rm neck}$, one might at first attribute the larger value of $\langle J_H\rangle$ obtained here compared with Marević et \textit{al.} to the different choice of the scission configuration. However, this interpretation is not supported by the present results. Indeed, for the configuration indicated by the first gray vertical line, characterized by $Q_{\rm neck}\approx0.7$, one obtains $\langle J_H\rangle=7.5\hbar$, which is a substantially larger value than the one reported by Marević et \textit{al.} for a configuration with $Q_{\rm neck}\approx2.5$, i.e., at a significantly larger neck density. The differences between the two calculations therefore cannot be attributed solely to the choice of the HFB configuration.

A more likely origin of this discrepancy is the different prescriptions used to define the fragments. In the procedure adopted by Marević et \textit{al.}, the spatial extension of each fragment is explicitly restricted, whereas the present separation procedure can retain a larger fraction of the low-density tails of the nuclear distributions within each fragment. These additional spatial contributions may increase the total AM assigned to the fragments. This interpretation is consistent with the systematic excess of the mean total angular momenta, $\langle J_L\rangle$ and $\langle J_H\rangle$, obtained here and suggests that the prescription used to separate the fragments represents an important source of the remaining differences between the two calculations. \\

\begin{figure}
\centering
\includegraphics[width=1.0\linewidth]{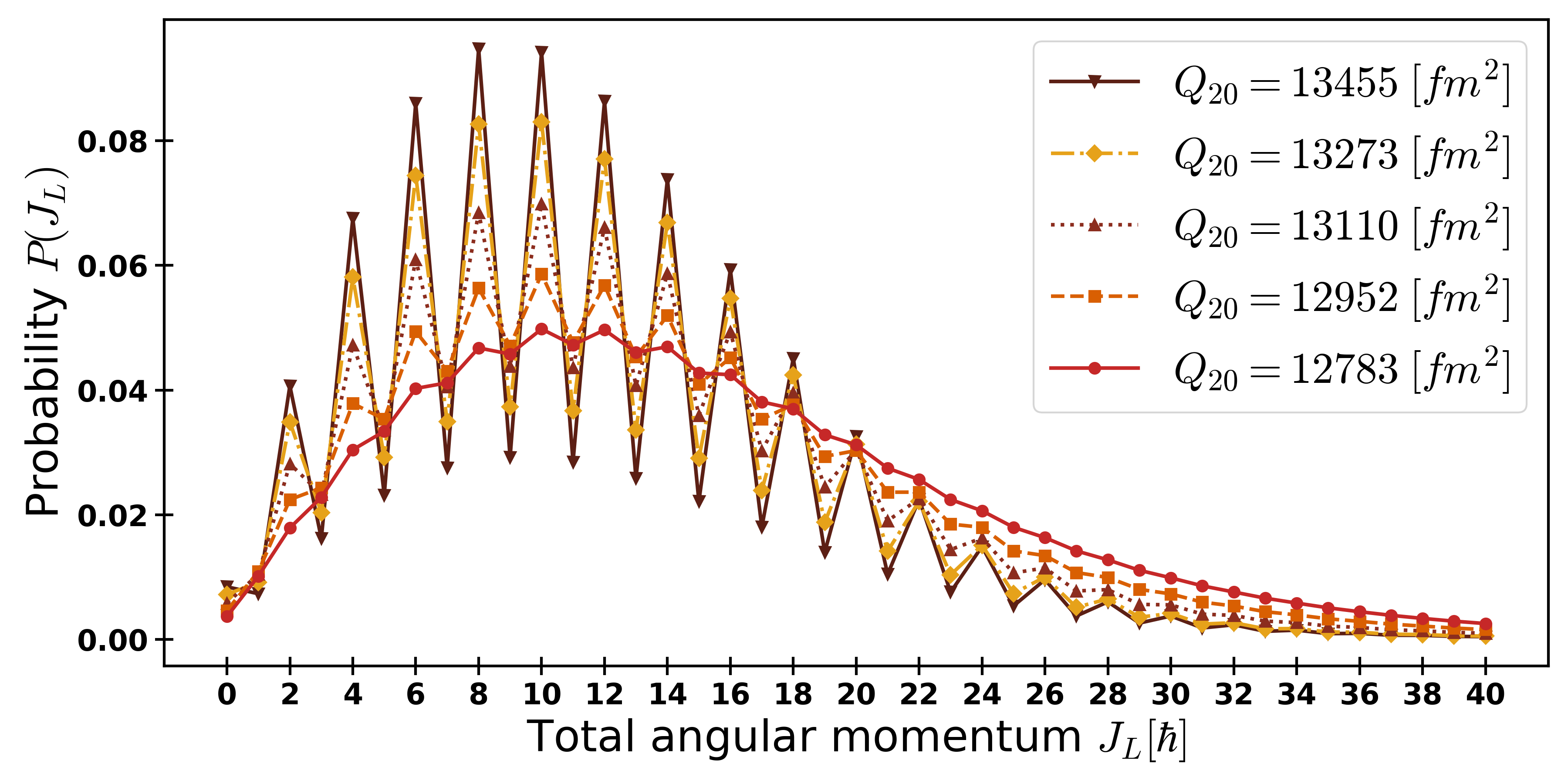}
\caption{Total angular momentum distributions $P(J_L)$ of the light fragment for different values of $Q_{20}$ in the vicinity of scission.}
\label{figVarL}
\end{figure}

\begin{figure}
\centering
\includegraphics[width=1.0\linewidth]{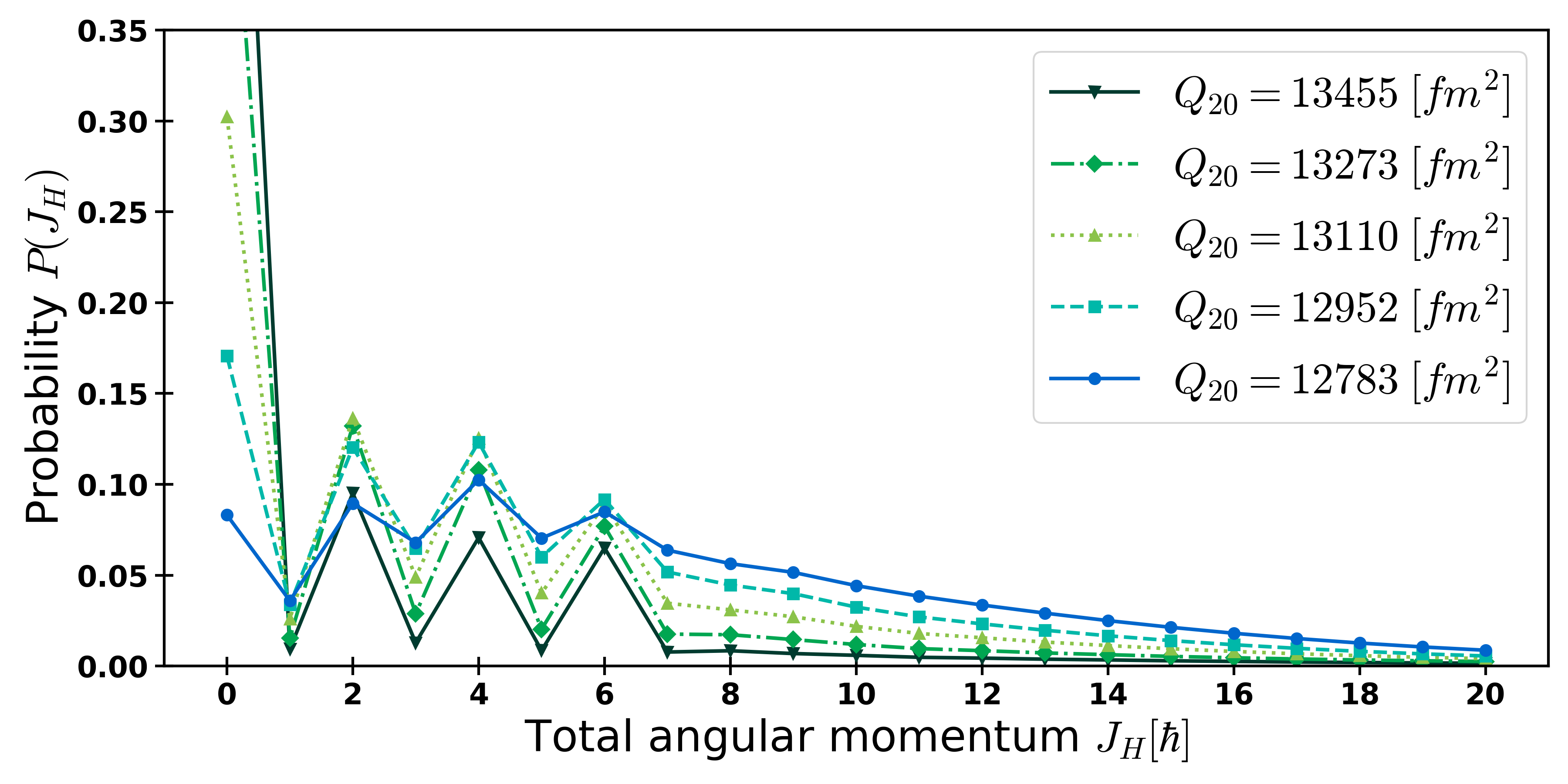}
\caption{Same as FIG. \ref{figVarL}, but for the heavy fragment.}
\label{figVarH}
\end{figure}

As a second step, to clarify the origin of the pronounced dominance of even-$J$ components observed in the present results but not in those of Marević et \textit{al.}, the total AM distributions obtained for different values of $Q_{20}$ are shown in FIGs. \ref{figVarL} and \ref{figVarH}. The two extreme configurations, characterized by $Q_{20}=12783~\mathrm{fm}^2$ and $Q_{20}=13455~\mathrm{fm}^2$, correspond to the two gray vertical lines in FIG. \ref{figIndic}, while the configuration at $Q_{20}=13110~\mathrm{fm}^2$ corresponds to the black vertical line, which also marks the peak of the chemical potential.
A clear increase in the dominance of even-$J$ components is observed with increasing $Q_{20}$, and therefore with decreasing $Q_{\rm neck}$. In the light fragment, the distribution becomes nearly smooth for $Q_{20}=12783~\mathrm{fm}^2$, with no pronounced even-$J$ dominance. These results therefore support the interpretation that the difference in the even-$J$ dominance observed between the present calculation and that of Marević et \textit{al.} is, to a significant extent, related to the different choice of the scission configuration.

To quantify and further elucidate this behavior, we introduce the odd-$J$ fraction $a(J_F)$. It is defined by:
\begin{equation}
    a(J_F) = \sum_{J_F\;{\rm odd}} P(J_F).
\end{equation}
This quantity directly measures the fraction of the AM distribution carried by odd values of $J$. In addition, the intrinsic quadrupole and octupole moments of each fragment, evaluated with respect to its center of mass, are considered. They are given by:
\begin{equation}
    Q_{20,F}^{CM} = Q_{20}^F - \sqrt{\frac{5}{4\pi}} A_F z_F^2,
\end{equation}
and
\begin{equation}
    Q_{30,F}^{CM} = \left|Q_{30}^F - \sqrt{\frac{63}{5}} z_F Q_{20}^F
    + \sqrt{\frac{7}{\pi}} A_F z_F^3\right|,
\end{equation}
where $z_F$ denotes the projected center-of-mass position of the fragment and the absolute value of $Q_{30,F}^{CM}$ is considered. 

\begin{figure}
\centering
\includegraphics[width=1.0\linewidth]{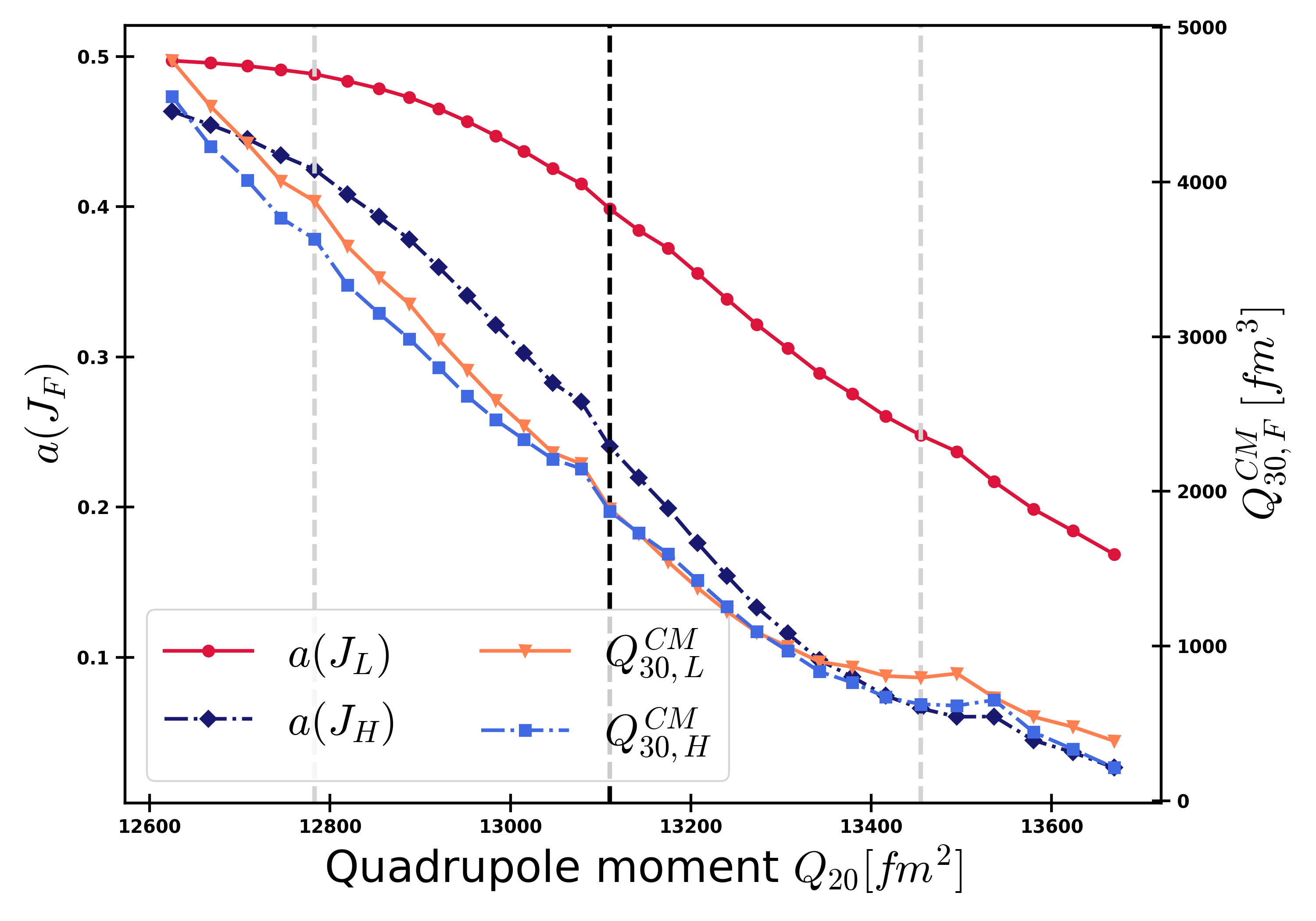}
\caption{Evolution of the odd-$J$ fraction $a(J_F)$ and the intrinsic octupole moment $Q_{30,F}^{CM}$ of the fragments as a function of $Q_{20}$.}
\label{figOdd}
\end{figure}

The evolution of $a(J_F)$ and $Q_{30,F}^{CM}$ with respect to $Q_{20}$ is shown in FIG. \ref{figOdd}.
A clear correlation is observed between the odd-$J$ fraction and the pronounced decrease of the intrinsic octupole moment in the vicinity of scission. As the octupole deformation of the fragments decreases, the odd-$J$ components are progressively suppressed, consistently with the restoration of reflection symmetry in the intrinsic fragment configurations. In the limit of a reflection-symmetric intrinsic state, the AM distribution is restricted to the $J$ values compatible with this symmetry, leading in the present $K=0$ case to the suppression of odd-$J$ components. The results therefore suggest that the increasing dominance of even-$J$ components originates from the rapid relaxation of the fragment deformations in the scission region, toward configurations that become increasingly reflection symmetric. 
This picture of rapid relaxation is further supported by the evolution of the intrinsic quadrupole moments, which are not shown in the figure. Over the same $Q_{20}$ interval, $Q_{20,L}^{CM}$ decreases from $758$ to $479~\mathrm{fm}^2$, while $Q_{20,H}^{CM}$ decreases from $247$ to $24~\mathrm{fm}^2$. 

The correlation between the odd-$J$ content and the rapid relaxation of the intrinsic fragment deformation suggests that the observed odd-even structure is unlikely to be coincidental. Rather, it may point to a genuine physical feature of the primary fragment total AM distributions at scission. This interpretation is further supported by recent results by Scamps et \textit{al.} \cite{SCAMPSLast}, obtained within a TDHFB framework, which also point toward a pronounced odd-even structure in the fragment total AM distributions. Our quasi-static analysis tends to suggest that the origin of this phenomenon come partly from geometrical deformation ones.\\

\begin{figure}
\centering
\includegraphics[width=1.0\linewidth]{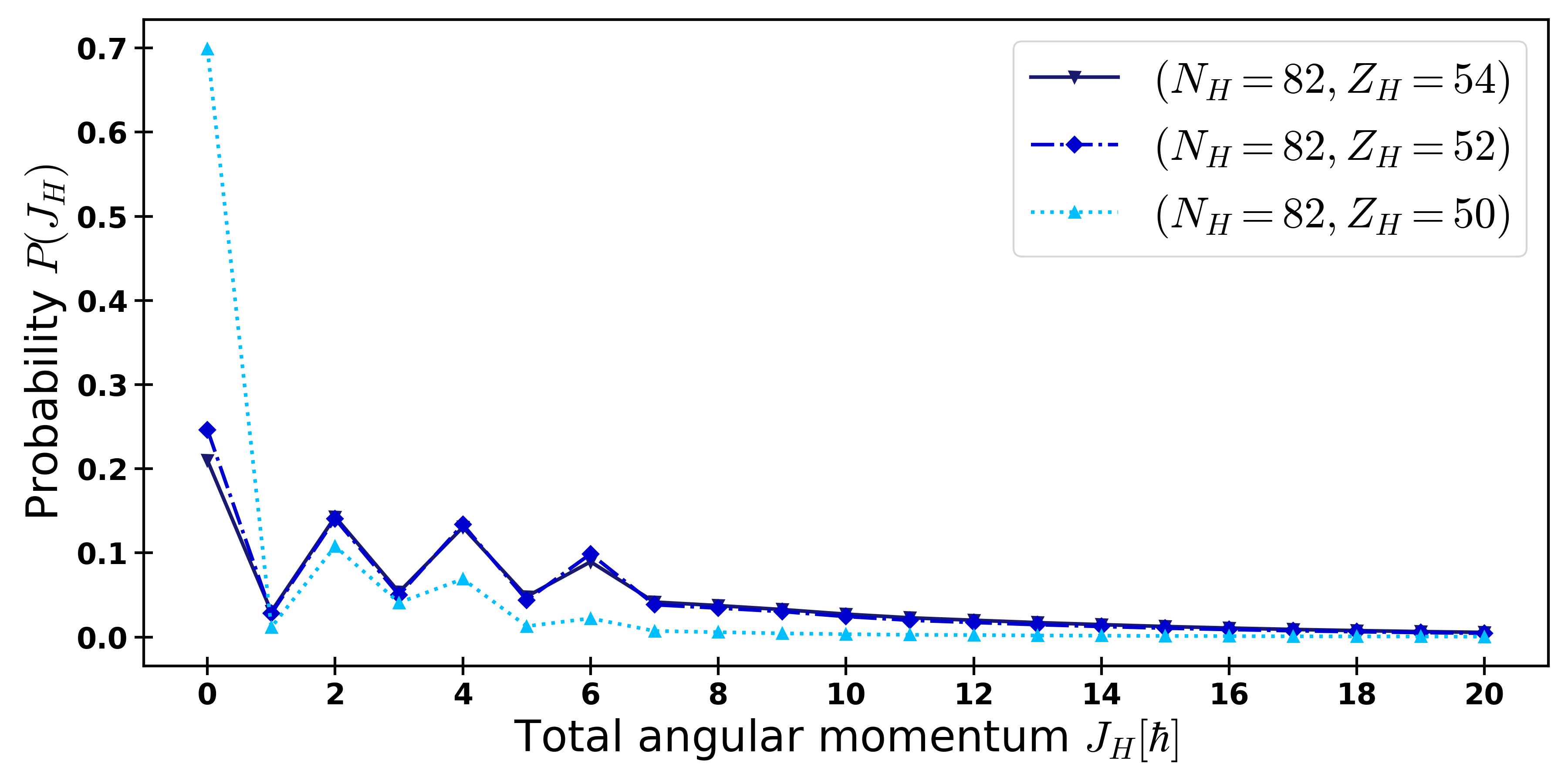}
\caption{Normalized total angular momentum distributions of the heavy fragment for the three dominant fragmentations at $Q_{20}=13110~\mathrm{fm}^2$. The corresponding probabilities in the HFB wave function are approximately $0.04$, $0.83$, and $0.13$ for $^{136}$Xe, $^{134}$Te, and $^{132}$Sn, respectively.}
\label{figPN}
\end{figure}

Finally, to investigate the origin of the large value of $P(J_H=0)$, which becomes even more pronounced in FIG. \ref{figVarH} as $Q_{20}$ increases and reaches a maximum value exceeding $0.6$, the normalized total AM distributions associated with the different significant fragmentations present in the HFB wave function at $Q_{20}=13110~\mathrm{fm}^2$ are shown separately in FIG. \ref{figPN}. In particular, the distributions corresponding to the fragmentations $(N_H=82,Z_H=54)$, $(N_H=82,Z_H=52)$, and $(N_H=82,Z_H=50)$ are considered. Their respective probabilities are approximately $0.04$, $0.83$, and $0.13$.
As seen in FIG. \ref{figPN}, the large value of $P(J_H=0)$ is primarily associated with the $(N_H=82,Z_H=50)$ fragmentation, corresponding to the doubly magic $^{132}$Sn nucleus. The strong shell effects associated with this doubly magic configuration may promote a more rapid relaxation of the fragment deformation toward a more spherical shape, potentially contributing to the enhancement of the $J_H=0$ component. This interpretation is supported by the intrinsic multipole moments listed in TABLE \ref{tabMom}, which indicate that the $^{132}$Sn fragment is substantially less deformed than the $^{136}$Xe and $^{134}$Te fragments, with both its quadrupole and octupole moments being strongly reduced.

\begin{table}[h!]
    \centering
    \renewcommand{\arraystretch}{1.25}
    \setlength{\tabcolsep}{10pt}
    \begin{tabular}{lccc}
        \toprule
        & $^{136}$Xe & $^{134}$Te & $^{132}$Sn \\
        \midrule
        $Q_{20,H}^{CM}\;[\mathrm{fm}^2]$
        & 141 & 124 & 42 \\[2pt]
        \addlinespace
        $Q_{30,H}^{CM}\;[\mathrm{fm}^3]$
        & 2409 & 2078 & 349 \\
        \bottomrule
    \end{tabular}
    \caption{Intrinsic quadrupole and octupole moments of the heavy fragments for the three dominant fragmentations at $Q_{20}=13110~\mathrm{fm}^2$.}
    \label{tabMom}
\end{table}

The absence of a similarly pronounced $J_H=0$ component in the results of Marević et \textit{al.} can then be understood as resulting from the combined effect of the deformation relaxation and the changing weights of the different fragmentations along the fission path. Indeed, the relaxation of the fragment deformation becomes particularly pronounced only at sufficiently small neck densities, while the probability of the $(N_H=82,Z_H=50)$ fragmentation increases significantly toward scission, from approximately $0.06$ at $Q_{20}=12783~\mathrm{fm}^2$ to $0.20$ at $Q_{20}=13455~\mathrm{fm}^2$. Thus, the increasing contribution of the $^{132}$Sn component, combined with its strong tendency toward a spherical shape, provides a natural explanation for the rapid increase of $P(J_H=0)$ as $Q_{20}$ increases. This behavior is consistent with the well-established role of shell effects in determining fragment properties and mass partitioning in nuclear fission, while suggesting that their influence also extends to the fine structure of the fragment total AM distributions. \\

\section{Conclusion}\label{section6}

In this article, we have proposed a new analytical method to extract the total AM distributions of fission fragments at scission within the HFB framework, using an underlying 2-center basis. The method relies on the generating functions of the HO wave functions, combined with Gaussian integration techniques and multivariate Hermite polynomials, to address the challenging coupling between the translations inherent to the 2-center basis and the rotations involved in the projection onto total AM. In addition, two theoretical simplifications with significant numerical consequences have been introduced. The first provides a straightforward treatment of the basis-completion problem arising from the rotation of the single-particle basis, while the second reduces the evaluation of the pfaffian to that of a simple determinant. These developments result in a formulation that is both analytically tractable and numerically efficient, providing a practical framework for the calculation of fission-fragment total AM distributions.

To benchmark the present approach against existing results, we have performed a first application for the $^{240}$Pu nucleus, a nucleus for which independent results are available in the literature \cite{marevic2025,SCAMPSLast}. More specifically, an asymmetric adiabatic fission path obtained using the recently proposed \enquote{Link} and \enquote{Drop} methods \cite{carpentier2024,TPaul,trilogy1} was considered. These methods provide a continuous description of the nucleus through scission and beyond. We have also adopted the canonical basis fragment separation procedure introduced in \cite{TPaul,trilogy1}, which is employed here for the first time in the context of total AM projection. This procedure provides a consistent definition of the fragment total AM operators in a basis that is in one-to-one correspondence with the single-particle basis in which the matrix elements of the single-particle rotation operator are calculated, and addresses the issues recently raised by L.M. Robledo \cite{robledo2026} concerning other total AM projection approaches.

Overall, the results obtained show a good qualitative agreement between the main features of the present approach and those reported by Marević et \textit{al.}~\cite{marevic2025}, in particular regarding the average total AM and the asymmetric sharing of total aAM between the light and heavy fragments. A systematic tendency toward larger total AM is nevertheless observed in the present study, which may be related to the prescription adopted for the spatial separation of the fragments.

The analysis of the differences between the two approaches has further revealed two features of particular interest. First, the strong suppression of the odd-$J$ components observed at large $Q_{20}$ values has been correlated with the rapid relaxation of the intrinsic fragment deformation in the vicinity of scission and with the restoration of a more nearly reflection-symmetric fragment shape. This correlation suggests that the resulting even-$J$ dominance may reflect a genuine physical feature of the primary fragment total AM distributions. Second, the pronounced contribution of the $^{132}$Sn component to the fine structure of the total AM distribution demonstrates that shell effects can influence not only the average fragment properties and mass partitioning, but also the fine structure of the fragment total AM distributions. These observations point to a common methodological requirement: sufficiently close-to-scission configurations must be considered in order to resolve the fine structure of the fragment total AM distributions.

The results presented here should nevertheless be regarded as a first illustration of the capabilities of the present approach. Further studies will be required to quantify the sensitivity of the predicted distributions to the fragment-separation prescription and to account for intrinsic excitations. \\

\noindent {\textbf{Acknowledgment}: P.C. and N.P. would like to thank D. Lacroix and P. Morfouace for valuable discussions on theoretical and experimental aspects, respectively. \\

\appendix

\section{Components of $I_G(\zeta)$}\label{appendixa}

\noindent The aim of this section is to provide the ingredients needed to express $I_G(\zeta)$ as the exponential of a quadratic form.

\subsection{Expressions for $M(\zeta)$ and $M^{-1}(\zeta)$}
\noindent The coefficients of the matrix $M(\zeta)$ are given by:
\begin{widetext}
\begin{eqnarray}
    &M(\zeta) = \begin{pmatrix}
        \frac{\displaystyle \text{cos}^2(\zeta)+1}{\displaystyle b_r^2} + \frac{\displaystyle \text{sin}^2(\zeta)}{\displaystyle b_z^2} & 0 & \text{cos}(\zeta)\text{sin}(\zeta)(\frac{\displaystyle 1}{\displaystyle b_z^2} - \frac{\displaystyle 1}{\displaystyle b_r^2}) \\
        0 & \frac{\displaystyle 2}{\displaystyle b_r^2} & 0 \\
         \text{cos}(\zeta)\text{sin}(\zeta)(\frac{\displaystyle 1}{\displaystyle b_z^2} - \frac{\displaystyle 1}{\displaystyle b_r^2}) & 0 &  \frac{\displaystyle \text{sin}^2(\zeta)}{\displaystyle b_r^2} + \frac{\displaystyle \text{cos}^2(\zeta)+1}{\displaystyle b_z^2}
    \end{pmatrix}.
\end{eqnarray}
\end{widetext}
\noindent In the following, the matrix elements of its inverse $M^{-1}(\zeta)$ will be noted as:
\begin{eqnarray}
     M^{-1}(\zeta)  &\overset{\text{def}}{=}& \begin{pmatrix}
        m_{11}(\zeta) & 0 & m_{13}(\zeta) \\
        0 & m_{22}(\zeta) & 0 \\ 
        m_{13}(\zeta) & 0 & m_{33}(\zeta)
    \end{pmatrix}.
\end{eqnarray}

\subsection{Expression for $J(\zeta)$}

\noindent Consistently with the partial derivatives, we introduce the transformations:
\begin{eqnarray}
    p_x &=& \frac{p_+ + p_-}{\sqrt{2}},
\end{eqnarray}
\begin{eqnarray}
    p_y &=& \frac{p_+-p_-}{i\sqrt{2}},
\end{eqnarray}
\begin{eqnarray}
   q_x &=& \frac{q_+ + q_-}{\sqrt{2}},
\end{eqnarray}
\begin{eqnarray}
     q_y &=& \frac{q_+-q_-}{i\sqrt{2}}.
\end{eqnarray}
\noindent One obtains for the expression of $J(\zeta)$ the following formula:
\begin{widetext}
\begin{eqnarray}
    J(\zeta)  &=& \begin{pmatrix}
        \displaystyle p_+J_{1p_+} + p_-J_{1p_-} + q_zJ_{1q_z}(\zeta)+ q_+J_{1q_+}(\zeta) + q_-J_{1q_-}(\zeta) + J_{1d'}(\zeta)\\  p_+J_{2p_+} + p_-J_{2p_-} + q_+J_{2q_+} + q_-J_{2q_-}\\ p_zJ_{3p_z} + q_zJ_{3q_z}(\zeta)+ q_+J_{3q_+}(\zeta) + q_-J_{3q_-}(\zeta) + J_{3dd'}(\zeta)
    \end{pmatrix},
\end{eqnarray}
\end{widetext}
where we have defined:
\begin{eqnarray}
    J_{1p_+} &=& \frac{\sqrt{2}}{b_r},
\end{eqnarray}
\begin{eqnarray}
    J_{1p_-} &=& \frac{\sqrt{2}}{b_r},
\end{eqnarray}
\begin{eqnarray}
    J_{1q_z}(\zeta) &=& \frac{2\text{sin}(\zeta)}{b_z}, 
\end{eqnarray}
\begin{eqnarray}
    J_{1q_+}(\zeta) &=& \frac{\sqrt{2}\text{cos}(\zeta)}{b_r},
\end{eqnarray}
\begin{eqnarray}
    J_{1q_-}(\zeta) &=& \frac{\sqrt{2}\text{cos}(\zeta)}{b_r}
\end{eqnarray}
\begin{eqnarray}
    J_{1d'}(\zeta) &=& -\frac{\text{sin}(\zeta)d'}{b_z^2},
\end{eqnarray}
\begin{eqnarray} 
    J_{2p_+} &=& \frac{\sqrt{2}}{ib_r}, 
\end{eqnarray}
\begin{eqnarray} 
    J_{2p_-} &=& -\frac{\sqrt{2}}{ib_r}, 
\end{eqnarray}
\begin{eqnarray} 
    J_{2q_+} &=& \frac{\sqrt{2}}{ib_r},
\end{eqnarray}
\begin{eqnarray} 
    J_{2q_-} &=& -\frac{\sqrt{2}}{ib_r}
\end{eqnarray}
\begin{eqnarray} 
    J_{3p_z} &=& \frac{2}{b_z},
\end{eqnarray}
\begin{eqnarray} 
    J_{3q_z}(\zeta) &=& \frac{2\text{cos}(\zeta)}{b_z}, 
\end{eqnarray}
\begin{eqnarray} 
    J_{3q_+}(\zeta) &=& -\frac{\sqrt{2}\text{sin}(\zeta)}{b_r}, 
\end{eqnarray}
\begin{eqnarray} 
     J_{3q_-}(\zeta) &=& -\frac{\sqrt{2}\text{sin}(\zeta)}{b_r},
\end{eqnarray}
\begin{eqnarray} 
     J_{3dd'}(\zeta) &=& -\frac{d+\text{cos}(\zeta)d'}{b_z^2}. 
\end{eqnarray}

\subsection{Expressions for $A(\zeta)$, $B(\zeta)$ and $C(\zeta)$}

\noindent The coefficients of the symmetric $6\times 6$ matrix $A(\zeta)$ are given by:
\begin{widetext}

\begin{eqnarray}
    A_{11}(\zeta) &=& -\frac{1}{2}m_{11}(\zeta)J_{1p_+}^2-\frac{1}{2}m_{22}(\zeta)J_{2p_+}^2, \\ A_{12}(\zeta) &=& -\frac{1}{2}\big(m_{11}(\zeta)J_{1p_+}J_{1p_-}+m_{22}(\zeta)J_{2p_+}J_{2p_-}\big) + 1, \\ A_{13}(\zeta) &=& -\frac{1}{2}m_{13}(\zeta)J_{1p_+}J_{3p_z},  \\ A_{14}(\zeta) &=& -\frac{1}{2}\big(m_{11}(\zeta)J_{1p_+}J_{1q_+}(\zeta) + m_{22}(\zeta)J_{2p_+}J_{2q_+} + m_{13}(\zeta)J_{1p_+}J_{3q_+}(\zeta)\big),
    \\ A_{15}(\zeta) &=& -\frac{1}{2}\big(m_{11}(\zeta)J_{1p_+}J_{1q_-}(\zeta) + m_{22}(\zeta)J_{2p_+}J_{2q_-} + m_{13}(\zeta)J_{1p_+}J_{3q_-}(\zeta)\big), \\ A_{16}(\zeta) &=& -\frac{1}{2}\big(m_{11}(\zeta)J_{1p_+}J_{1q_z}(\zeta) + m_{13}(\zeta)J_{1p_+}J_{3q_z}(\zeta)\big)
\end{eqnarray}

\begin{eqnarray}
    A_{22}(\zeta) &=& -\frac{1}{2}m_{11}(\zeta)J_{1p_-}^2-\frac{1}{2}m_{22}(\zeta)J_{2p_-}^2,  \\ A_{23}(\zeta) &=& -\frac{1}{2}m_{13}(\zeta)J_{1p_-}J_{3p_z}, \\
    A_{24}(\zeta) &=& -\frac{1}{2}\big(m_{11}(\zeta)J_{1p_-}J_{1q_+}(\zeta) + m_{22}(\zeta)J_{2p_-}J_{2q_+} + m_{13}(\zeta)J_{1p_-}J_{3q_+}(\zeta)\big), \\ A_{25}(\zeta) &=& -\frac{1}{2}\big(m_{11}(\zeta)J_{1p_-}J_{1q_-}(\zeta) + m_{22}(\zeta)J_{2p_-}J_{2q_-} + m_{13}J_{1p_-}J_{3q_-}(\zeta)\big), \\
    A_{26}(\zeta) &=& -\frac{1}{2}\big(m_{11}(\zeta)J_{1p_-}J_{1q_z}(\zeta) + m_{13}(\zeta)J_{1p_-}J_{3q_z}(\zeta)\big),
\end{eqnarray}

\begin{eqnarray}
    A_{33}(\zeta) &=& -\frac{1}{2}m_{33}(\zeta)J_{3p_z}^2 + 1, \\ A_{34}(\zeta) &=& -\frac{1}{2}\big(m_{33}(\zeta)J_{3p_z}J_{3q_+}(\zeta)+m_{13}(\zeta)J_{1q_+}(\zeta)J_{3p_z}\big), \\
    A_{35}(\zeta) &=& -\frac{1}{2}\big(m_{33}(\zeta)J_{3p_z}J_{3q_-}(\zeta)+m_{13}(\zeta)J_{1q_-}(\zeta)J_{3p_z}\big), \\ A_{36}(\zeta) &=& -\frac{1}{2}\big(m_{33}(\zeta)J_{3p_z}J_{3q_z}(\zeta)+m_{13}(\zeta)J_{1q_z}(\zeta)J_{3p_z}\big),
\end{eqnarray}

\begin{eqnarray}
    A_{44}(\zeta) &=& -\frac{1}{2}m_{11}(\zeta)J_{1q_+}^2(\zeta)-\frac{1}{2}m_{22}(\zeta)J_{2q_+}^2 -\frac{1}{2}m_{33}(\zeta)J_{3q_+}^2(\zeta) -m_{13}(\zeta)J_{1q_+}(\zeta)J_{3q_+}(\zeta), \\ A_{45}(\zeta) &=& -\frac{1}{2}\big(m_{11}(\zeta)J_{1q_+}(\zeta)J_{1q_-}(\zeta)+m_{22}(\zeta)J_{2q_+}J_{2q_-}+m_{33}(\zeta)J_{3q_+}(\zeta)J_{3q_-}(\zeta) \\ && \qquad \qquad \qquad + m_{13}(\zeta)(J_{1q_+}(\zeta)J_{3q_-}(\zeta)+J_{1q_-}(\zeta)J_{3q_+}(\zeta))\big) + 1, \nonumber \\ 
    A_{46}(\zeta) &=& -\frac{1}{2}\big(m_{11}(\zeta)J_{1q_+}(\zeta)J_{1q_z}(\zeta)+m_{33}(\zeta)J_{3q_+}(\zeta)J_{3q_z}(\zeta)+m_{13}(\zeta)(J_{1q_+}(\zeta)J_{3q_z}(\zeta)+J_{1q_z}(\zeta)J_{3q_+}(\zeta))\big),
\end{eqnarray}

\begin{eqnarray}
    A_{55}(\zeta) &=& -\frac{1}{2}m_{11}(\zeta)J_{1q_-}^2(\zeta)-\frac{1}{2}m_{22}(\zeta)J_{2q_-}^2 -\frac{1}{2}m_{33(\zeta)}J_{3q_-}^2(\zeta) -m_{13}(\zeta)J_{1q_-}(\zeta)J_{3q_-}(\zeta), \\
    A_{56}(\zeta) &=& -\frac{1}{2}\big(m_{11}(\zeta)J_{1q_-}(\zeta)J_{1q_z}(\zeta)+m_{33}(\zeta)J_{3q_-}(\zeta)J_{3q_z}(\zeta)+m_{13}(\zeta)(J_{1q_-}(\zeta)J_{3q_z}(\zeta)+J_{1q_z}(\zeta)J_{3q_-}(\zeta))\big), 
\end{eqnarray}

\begin{eqnarray}
    A_{66}(\zeta) &=& -\frac{1}{2}m_{11}(\zeta)J_{1q_z}^2(\zeta) -\frac{1}{2}m_{33}(\zeta)J_{3q_z}^2(\zeta) -m_{13}(\zeta)J_{1q_z}(\zeta)J_{3q_z}(\zeta) + 1.
\end{eqnarray}

\noindent The coefficients of the six-dimensional vector $B(\zeta)$ are then given by:
\begin{eqnarray}
    B_1(\zeta) &=& m_{11}(\zeta)J_{1p_+}J_{1d'}(\zeta) + m_{13}(\zeta)J_{1p_+}J_{3dd'}(\zeta), \\
    B_2(\zeta) &=& m_{11}(\zeta)J_{1p_-}J_{1d'}(\zeta) + m_{13}(\zeta)J_{1p_-}J_{3dd'}(\zeta), \\
    B_3(\zeta) &=& m_{33}(\zeta)J_{3p_z}J_{3dd'}(\zeta) + m_{13}(\zeta)J_{1d'}(\zeta)J_{3p_z} + \frac{2d}{b_z}, \\
    B_4(\zeta) &=& m_{11}(\zeta)J_{1q_+}(\zeta)J_{1d'}(\zeta) + m_{33}(\zeta)J_{3q_+}(\zeta)J_{3dd'}(\zeta)+m_{13}(\zeta)(J_{1q_+}(\zeta)J_{3dd'}(\zeta)+J_{1d'}(\zeta)J_{3q_+}(\zeta)), \\
    B_5(\zeta) &=& m_{11}(\zeta)J_{1q_-}(\zeta)J_{1d'}(\zeta) + m_{33}(\zeta)J_{3q_-}(\zeta)J_{3dd'}(\zeta)+m_{13}(\zeta)(J_{1q_-}(\zeta)J_{3dd'}(\zeta)+J_{1d'}(\zeta)J_{3q_-}(\zeta)), \\
    B_6(\zeta) &=& m_{11}(\zeta)J_{1q_z}(\zeta)J_{1d'}(\zeta) + m_{33}(\zeta)J_{3q_z}(\zeta)J_{3dd'}(\zeta) + m_{13}(\zeta)(J_{1q_z}(\zeta)J_{3dd'}(\zeta)+J_{1d'}(\zeta)J_{3q_z}(\zeta)) + \frac{\displaystyle 2d'}{\displaystyle b_z}.
\end{eqnarray}
\noindent Finally, the scalar $C(\zeta)$ is given by:
\begin{eqnarray}
    C(\zeta) = \frac{1}{2}m_{11}(\zeta)J_{1d'}^2(\zeta) +\frac{1}{2}m_{33}(\zeta)J_{3dd'}^2(\zeta) +m_{13}(\zeta)J_{1d'}(\zeta)J_{3dd'}(\zeta).
\end{eqnarray}
\end{widetext}

\bibstyle{apsrev4-2}

\bibliography{main}

\end{document}